\documentclass[letterpaper]{nature}
\usepackage[labelfont=bf]{caption}  
\usepackage{subcaption}
\usepackage{float}
\usepackage{array,multirow}
\usepackage{booktabs}
\usepackage{nameref}
\usepackage{xcolor,colortbl} 
\usepackage{todonotes}
\usepackage{enumitem}
\usepackage{amsfonts}
\usepackage{amssymb}
\usepackage{mathbbol}
\usepackage{siunitx}
\usepackage{authblk}
\usepackage{amsmath}
\usepackage{graphicx}
\usepackage{lineno}

\definecolor{Gray}{gray}{0.85}

\makeatletter
\let\saved@includegraphics\includegraphics
\AtBeginDocument{\let\includegraphics\saved@includegraphics}

\renewenvironment{figure}
  {\@float{figure}\centering}
  {\end@float}
\makeatother

\usepackage[numbers,sort&compress]{natbib}

\usepackage{hyperref}
\hypersetup{
    colorlinks=true,
    linkcolor=blue,
    citecolor=blue,
    urlcolor=blue
}

\def\Bb   {{\bf{B}}}

\def\Eb   {{\bf{E}}}
\def\Hb   {{\bf{H}}}
\def\Heff {{\bf{H}_{\rm eff}}}
\def\Jb   {{\bf{J}}}
\def\Mb   {{\bf{M}}}

\newcommand{\ii}{\mathrm{i}}

\title{HPC-Driven Modeling with ML-Based Surrogates for Magnon-Photon Dynamics in Hybrid Quantum Systems}

\author[1,2]{Jialin Song}
\author[3]{Yingheng Tang}
\author[2]{Pu Ren}
\author[4]{Shintaro Takayoshi}
\author[3]{Saurabh Sawant}
\author[5]{Yujie Zhu}
\author[5]{Jia-Mian Hu}
\author[3]{Andy Nonaka}
\author[2,6,7]{Michael W. Mahoney}
\author[2,6]{Benjamin Erichson\textsuperscript{*}}
\author[3]{Zhi (Jackie) Yao\textsuperscript{*}}

\affil[1]{Simon Fraser University, 8888 University Dr W, Burnaby, BC V5A 1S6, Canada}
\affil[2]{Scientific Data Division, Lawrence Berkeley National Laboratory, CA, 94720, USA}
\affil[3]{Applied Mathematics and Computational Research Division, Lawrence Berkeley National Laboratory, CA, 94720, USA}
\affil[4]{Konan University, Japan}
\affil[5]{Department of Materials Science and Engineering, University of Wisconsin-Madison, Madison, Wisconsin 53706, USA}
\affil[6]{International Computer Science Institute, Berkeley, CA 94704, USA}
\affil[7]{Department of Statistics, University of California, Berkeley, CA 94720, USA}

\graphicspath{{figures/}}
\date{}
\begin{document}

\maketitle

\begingroup
\renewcommand\thefootnote{\fnsymbol{footnote}} 
\footnotetext[1]{Email: \href{mailto:jackie_zhiyao@lbl.gov}{\texttt{jackie\_zhiyao@lbl.gov}}, \href{mailto:erichson@lbl.gov}{\texttt{erichson@lbl.gov}}}
\endgroup


\vspace{0.5em}
\begin{abstract}
    Simulating hybrid magnonic quantum systems remains a challenge due to the large disparity between the timescales of the two systems. 
    We present a massively parallel GPU-based simulation framework that enables fully coupled, large-scale modeling of on-chip magnon-photon circuits. 
    Our approach resolves the dynamic interaction between ferromagnetic and electromagnetic fields with high spatiotemporal fidelity. 
    To accelerate design workflows, we develop a physics-informed machine learning surrogate trained on the simulation data, reducing computational cost while maintaining accuracy. 
    This combined approach reveals real-time energy exchange dynamics and reproduces key phenomena such as anti-crossing behavior and the suppression of ferromagnetic resonance under strong electromagnetic fields. 
    By addressing the multiscale and multiphysics challenges in magnon–photon modeling, our framework enables scalable simulation and rapid prototyping of next-generation quantum and spintronic devices.
\end{abstract}

\section{Introduction}\label{sec:introduction}
Hybrid quantum systems, which combine distinct physical platforms, are a promising route toward advanced quantum technologies, as they harness strong interactions that may not be readily achievable in a single platform~\cite{Nielsen_Chuang_2010, 2015Gershon}.  
These systems take many forms, coupling any two (or more) quantum platforms — for example, superconducting qubits ~\cite{Scarlino2019, Tabuchi2015}, microwave resonators ~\cite{Clerk2020}, single spins~\cite{Viennot2015}, spin ensembles ~\cite{Tabuchi2015, Hou2019, Li2019, Chen2018}, or mechanical resonators~\cite{o2010phonon, andrews2015phonon, manenti2017phonon} — to harness strong interactions.
These heterogeneous systems leverage complementary advantages of each component, but their rich multi-physics interactions pose formidable modeling challenges. 
A prominent example is cavity magnonics, where collective spin excitations (magnons) couple with microwave photons in a resonant cavity to form hybrid magnon–polariton modes when tuned into resonance~\cite{Zhang2014, ZareRameshti2022, Awschalom2021}. 
While an avoided crossing indicates hybridization, it is not sufficient for strong coupling; the system enters the strong-coupling regime only when the coupling rate $g$ (half the on‑resonance splitting) exceeds the dissipation rates of the uncoupled magnon and photon modes, in which case energy oscillates coherently between them and the hybrid modes exhibit remarkable coherence owing to the high $Q$ factors of microwave resonators and the narrow ferromagnetic resonance (FMR) linewidth~\cite{Match2019}.
These states are essential for quantum operations such as mode swapping~\cite{Match2019, zhuang2024dynamical}, quantum state storage ~\cite{Zhang2015, Tabuchi2015, Pishehvar2024}, and dynamic control of energy exchange ~\cite{Xu2020, Pishehvar2024}.
The hallmark experimental signature of strong magnon--photon coupling is a pronounced avoided crossing (mode splitting) in the frequency spectrum, in agreement with theoretical predictions~\cite{MagnonPhotonTheory} and observed in many 3D ~\cite{Zhang2014, Goryachev2014} and on-chip 2D~\cite{Huebl2013, Hou2019, Li2019} cavity based systems. 
In the time domain, the coupled system exhibits Rabi-like oscillations with very high energy exchange per cycle, reflecting the complete hybridization of the two excitations into a single quantum system~\cite{zhuang2024dynamical}.
Achieving practical quantum operations with magnon-photon systems requires precise dynamical control over energy transfer between the two hybrid magnon-polariton modes~\cite{Boventer2020, Golovchanskiy2021}. 
Therefore, developing quantum devices that leverage these effects necessitates a detailed understanding of their time-domain dynamics.

However, accurately modeling the complex dynamics of such hybrid systems presents a significant computational challenge. 
Existing modeling approaches have been limited in scope. 
Most analytical treatments require inputting an assumed magnon--photon coupling strength, and the few numerical studies to date have often resorted to reduced dimensions (1D or 2D models) or macrospin approximations to mitigate computational cost~\cite{MagnonPhotonTheory}. 
Conventional micromagnetic modeling strategies rely on the magnetostatic approximation~\cite{donahue1999oommf, mumax3, bjork2021magtense}, which has provided valuable insights, but which typically neglects the dynamic nature of electromagnetic (EM) fields, including time-dependent displacement currents. 
The underlying physics of magnon-photon coupling spans multiple domains and scales, requiring a multiphysics simulation that solves Maxwell's equations for electromagnetic fields alongside the Landau--Lifshitz--Gilbert (LLG) equation for magnetization dynamics~\cite{zhuang2024dynamical}. 
Solving these coupled equations self-consistently in three dimensions with realistic material parameters is computationally intensive, often necessitating high-performance computing (HPC) resources to achieve adequate spatial and temporal resolution~\cite{Yao2022}. 
The disparate time scales and length scales involved -- from microwave photons to nanometer-scale spin precessions -- further complicate direct simulation.
Maxwell-LLG coupling is achieved by self-consistently updating the dynamic EM fields in sync with magnetization dynamics, thereby capturing the bidirectional interaction between microwave photons and magnons ~\cite{Yao2022, Yao2021, Yao2018, zhuang2024dynamical, xu2024slow, pishehvar2025demand, maksymov2015rigorous, yu2020circulating}.
The majority of such time-domain dynamical simulations employ either explicit schemes like finite-difference time-domain (FDTD) leapfrog integrators ~\cite{Yao2022, maksymov2015rigorous, zhuang2024dynamical, pishehvar2025demand, xu2024slow}, or implicit schemes such as alternating-direction implicit (ADI) FDTD and Crank–Nicolson methods ~\cite{Yao2021, Yao2018}. 
While explicit FDTD is straightforward, it is only conditionally stable (restricted by the Courant limit), leading to significant oversampling of the EM wave in the time domain. 
To ease numerical constraints, researchers have explored strategies such as artificially increasing the medium permittivity—an approach that relaxes time-step oversampling~\cite{zhuang2024dynamical}. 
However, this strategy might be difficult to extend to complex device structures with highly inhomogeneous local permittivity and conductivities, as in on-chip devices.  
On the other hand, implicit methods offer improved (often unconditional) stability, but they require solving more complex update steps.
Additionally, implementing precise boundary conditions in implicit mathematical frameworks, such as perfectly matched layers (PML), is challenging and limits scalability in real-world quantum devices ~\cite{Yao2018}.
HPC techniques, including GPU acceleration and parallel FDTD-LLG solvers combined with domain decomposition, significantly enhance simulation speed and scalability, enabling large-scale modeling on modern supercomputing architectures ~\cite{fu2015finite, Yao2022}.
Finite-element time-domain models that rely on commercial software have also been adopted, but they suffer from the overhead of assembling and solving large linear systems at each time step, leading to high CPU and memory usage \cite{yu2020circulating}.
Nonetheless, to the best of the authors' knowledge, dynamical time-domain simulation of practical sized, on-chip hybrid magnon–photon circuits has remained elusive~\cite{Hou2019, Li2019}.
This underscores the need for more powerful computational methods to simulate hybrid magnon--photon systems in full detail. 

The main numerical challenges arise from the need for large-scale three-dimensional (3D) simulations and fine meshes to accurately capture the detailed structures.
In many practical scenarios, however, only a few spatial observation points are needed for circuit analysis tasks. 
This discrepancy underscores the potential for more efficient modeling strategies that focus resources on critical regions and time windows, reducing overall costs while preserving key physics.
Machine learning (ML) has emerged as a powerful approach to tackle the computational challenges posed by high-dimensional and nonlinear dynamical systems~\cite{brunton2020machine,wang2023scientific}. 
In particular, frameworks such as 
Physics-Informed Neural Networks (PINNs)~\cite{karniadakis2021physics,raissi2019physics}, Neural Operators (e.g., DeepONet ~\cite{lu2021learning} and Fourier Neural Operator~\cite{li2020fourier}), and 
Neural Discrete Equilibrium (NeurDE)~\cite{neurde_TR}
have demonstrated strong capabilities in modeling partial differential equations (PDEs), by efficiently capturing their underlying spatial-temporal dynamics. 
Furthermore, recent advances in sequence modeling architectures, including Recurrent Neural Networks (RNNs)~\cite{hochreiter1997long}, state-space models (SSMs)~\cite{gu2022s4_iclr,annan_tuning_TR} and Transformers~\cite{vaswani2017attention}, have shown remarkable success in predicting complex time-evolving physical phenomena~\cite{Tang2022,Ravuri2021,rao2023encoding,bi2023accurate,lam2023learning,song2024forecasting,lyu2024wavecastnet,learnConservation1_TR}. 

Despite significant advancements, accurately predicting long-horizon and multi-scale dynamics remains a challenge, particularly in coupled physical systems such as EM fields. Furthermore, classic data-driven ML approaches often require vast amounts of training data and struggle with generalization to out-of-distribution (OOD) scenarios, limiting their applicability in complex scientific domains. 

In this work, we present a hybrid computational framework that combines HPC-driven numerical modeling with physics-informed ML-based surrogates to address these limitations. By combining short-duration high-fidelity numerical simulations with these data-driven models, one can learn the latent electric and magnetic field patterns, enabling reliable predictions of magnon-photon interactions across the entire spatial and temporal domain. 
This hybrid strategy drastically cuts down the need for computationally expensive numerical solvers to a limited time window, while using ML to extrapolate the remaining dynamics with embedded physics knowledge. 
Consequently, researchers can achieve significant speedups without sacrificing precision or scalability when exploring complex coupled phenomena. 
Specifically, our main contributions include the following.
\begin{itemize}
    \item We present an HPC-enabled FDTD framework for large-scale 3D simulations, directly integrating Maxwell's equations with the LLG equation to dynamically model magnon-photon interactions. This framework integrates short-duration simulation data with ML methods to enable fast and accurate surrogate modeling of long-term and multi-scale EM systems. 
    \vspace{-0.3cm}
    \item We incorporate a domain-specific physics constraint into this hybrid ML framework to enhance training efficiency and ensure the preservation of fundamental physical principles, ultimately improving the OOD generalization capability. 
    \vspace{-0.3cm}
    \item We have evaluated the performance of our proposed method across extensive numerical experiments. The results demonstrate the effectiveness of our hybrid method in terms of simulation accuracy and computational efficiency. 
\end{itemize}

By incorporating a physics-informed ML network, our framework achieves highly accurate predictions of time-domain resonance responses using input sequences as short as $20\%$ of their full length. This advancement enables up to 5$\times$ acceleration, compared to standalone HPC simulations, representing a significant leap in both computational efficiency and predictive precision.
Crucially, this hybrid modeling strategy combines first-principles physics-based simulation with opportunities for data-driven enhancement, aligning with the emerging paradigm of \textit{scientific machine learning} in computational science. 
For example, the rich output data from our high-fidelity simulations could be used to train surrogate models or inform ML-based optimization routines, illustrating how data-driven methods can complement and accelerate traditional simulations. 
By uniting advanced multiphysics simulation with such data-driven techniques, our work showcases how innovations in computational science -- from HPC-driven algorithms to modern ML -- can broaden the exploration of complex quantum systems. 
Using the magnon--photon platform as a representative case, we demonstrate a cross-disciplinary approach that not only advances the modeling of a specific physical system but also exemplifies general strategies for tackling grand challenges in hybrid quantum computing and simulation.

\section{Results}
\label{sec:results}

\subsection{Magnon-Photon Hybrid Quantum Circuit Modeling Setup.}
\label{subsec:setup}

\begin{figure}
    \centering
    \includegraphics[width=\linewidth]{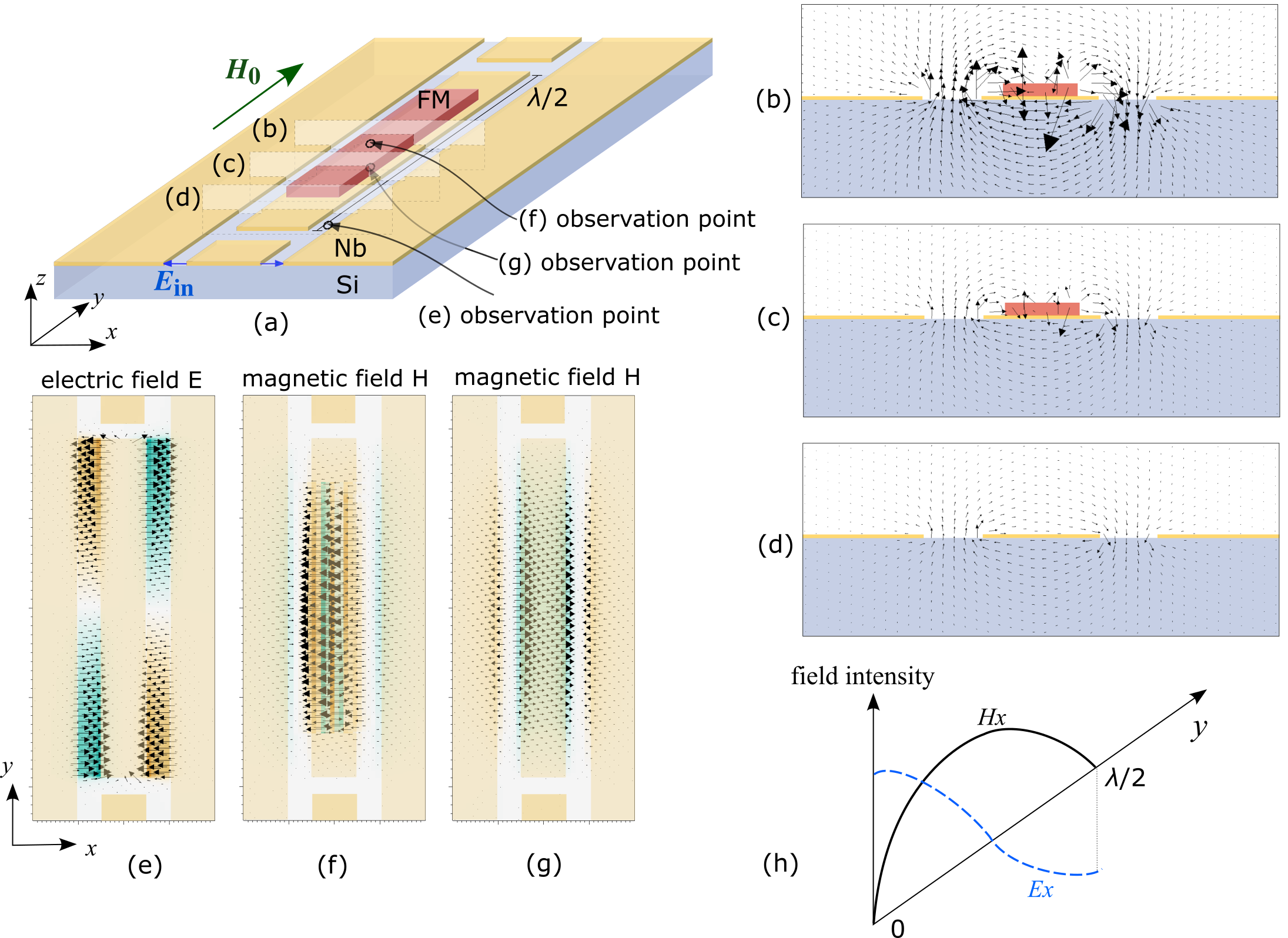}
    \caption{\textbf{Magnon-photon hybrid quantum circuit modeling setup and simulated field profiles.}
    (a) Perspective view of the magnon-photon coupling circuit, comprising a CPW resonator and a ferromagnetic thin film. The CPW resonator consists of a center conductor and lateral ground conductors (yellow) positioned on a silicon substrate (blue). The ferromagnet, shown as a red stripe, is placed on top of the center conductor. Material and geometrical parameters are detailed in the main text.
    (b)–(d) Cross-sectional views of the circuit, overlaid with the simulated magnetic field (\Hb) distribution at different locations along the longitudinal direction of the CPW resonator, as indicated in (a).
    (e)–(g) Top views of the circuit, overlaid with the simulated electric and magnetic field distributions in the z-plane at different observation points, as marked in (a).
    (h) Theoretical field distributions along the longitudinal direction of the circuit, where the electric field $E_x$ is concentrated in the lateral air gap, and the magnetic field $H_x$ is observed above and beneath the center conductor. }
    \label{fig:fig1-overview}
\end{figure}

\noindent
The simulation setup is shown in Figure~\ref{fig:fig1-overview}(a), where the core component is an on-chip coplanar waveguide (CPW) microwave resonator designed to generate and support photons. 
The metal conductivity is $1.0\times 10^{11}\,\mathrm{S/m}$ to minimize resistive losses and approximate superconducting behavior, and the relative permittivity of the silicon substrate is $\epsilon_\mathrm{Si}=11.4$.
The metal layer is $2\,\mathrm{\mu m}$ thick, and the silicon substrate is $500\,\mathrm{\mu m}$ thick.
The center line width is $50\,\mathrm{\mu m}$, and each lateral gap between the center line and the adjacent ground patches is $25\,\mathrm{\mu m}$. 
Such geometrical and material design results in a characteristic impedance of 50 $\Omega$ for the CPW transmission line.
This CPW resonator is tuned to a fundamental mode at $f_\textrm{p} = \omega_\textrm{p}/2\pi = 15.2\,\mathrm{GHz}$, with a center conductor line length of $3.75\,\mathrm{mm}$. 
At both ends of the resonator (along $\pm y$), feeding metallic patches of $300\,\mathrm{\mu m}$ in length and $50\,\mathrm{\mu m}$ in width are implemented. 
Gap capacitors are formed by $200\,\mathrm{\mu m}$ gaps between the central conductor and the feeding conductors. 

A magnetic thin film (red stripe in Figure~\ref{fig:fig1-overview}(a)) is placed at the center of the conductor line. 
This film is $5\,\mathrm{\mu m}$ thick, $30\,\mathrm{\mu m}$ wide, occupying the center portion of the conductor, and $2.82\,\mathrm{mm}$ long, covering about three quarters of the resonator length. 
The saturation magnetization of the material is $4\pi M_S = 1.2\times 10^4\,\mathrm{Gauss}$ (typical of Permalloy), and the Gilbert damping factor is $\alpha=0.003$, indicating low magnetic loss. 
The electrical conductivity of the magnetic material is set to zero to avoid internal electrical losses, confining losses to the CPW resonator.
A DC magnetic bias field $H_0$ is applied in the longitunidal direction, to saturate the ferromagnet to its saturation magnetiation $M_S$. 
The value of $H_0$ ranges from 1500 Oersted to 2600 Oersted. 
All six boundaries of the simulation domain are set as PMLs. 
The top PML boundary in the $z$ direction is placed $1\,\mathrm{mm}$ above the bottom, leaving $493\,\mathrm{\mu m}$ of vacuum above the magnetic thin film surface. 
The bottom PML boundary is applied to the lower surface of the silicon substrate, effectively extending the substrate thickness beyond its physical size. 
Similarly, the lateral metal ground planes (width $250\,\mathrm{\mu m}$) also have PML boundary conditions in the $x$ direction, extending both the ground planes and the substrate. 
The longitudinal PML boundaries in the $y$ direction also effectively extend the feeding patches and the associated substrate infinitely, eliminating reflections from both feeding patches.
To excite magnon-photon polaritons in this structure, a voltage is applied at one end of the CPW from its center conductor to the lateral grounds (see Figure~\ref{fig:fig1-overview}(a)). The voltage excitation follows a modified Gaussian pulse, given by 
\( E_\mathrm{in} = \exp\left(-(t - 3T_p)^2/2T_p^2 \right) \cos(\omega_0 t) \), 
where \( \omega_0 = 2\pi f_0 \) centers the frequency at \( 16\,\mathrm{GHz} \), and \( T_p = 0.0625\,\mathrm{ns} \) represents one period of the excitation.
The simulation uses mesh sizes of $\Delta x\times\Delta y\times\Delta z=5\,\mathrm{\mu m}\times19\,\mathrm{\mu m}\times1\,\mathrm{\mu m}$, and the domain dimensions are $ x\times y\times z=0.6\,\mathrm{mm}\times4.75\,\mathrm{mm} \times1\,\mathrm{mm}$, yielding a grid of $N_x\times N_y\times N_z=120\times250\times1000$. 
A CFL factor of $0.9$ sets the time step to $\Delta t=2.94\times10^{-15}\,\mathrm{s}$.

In the fundamental mode, the CPW resonator has electric-field peaks at both ends of the center conductor and a magnetic-field peak in the middle (Figure~\ref{fig:fig1-overview}(h)). Placing the ferromagnet on top disturbs the magnetic field \Hb, creating an irregular vortex in regions where the ferromagnet is present (Figures~\ref{fig:fig1-overview}(b)\textendash(c)), while regions without ferromagnet retain the regular vortex (Figure~\ref{fig:fig1-overview}(d)). Figures~\ref{fig:fig1-overview}(f)\textendash(g) further show that the \Hb \hspace{0.25em} fields above and below the center metal point in opposite directions, consistent with Figures~\ref{fig:fig1-overview}(b)\textendash(d). In contrast, the ferromagnet minimally affects the electric field \Eb, which primarily spans from the center conductor to the ground and forms a standard standing-wave pattern with peaks at the ends and a null at the center (Figure~\ref{fig:fig1-overview}(e)).

\subsection{Fully-Coupled Dynamic Numerical Framework and Physics-Informed ML Surrogates.}
\label{subsec:models}

\begin{figure}
    \centering
    \includegraphics[width=\linewidth]{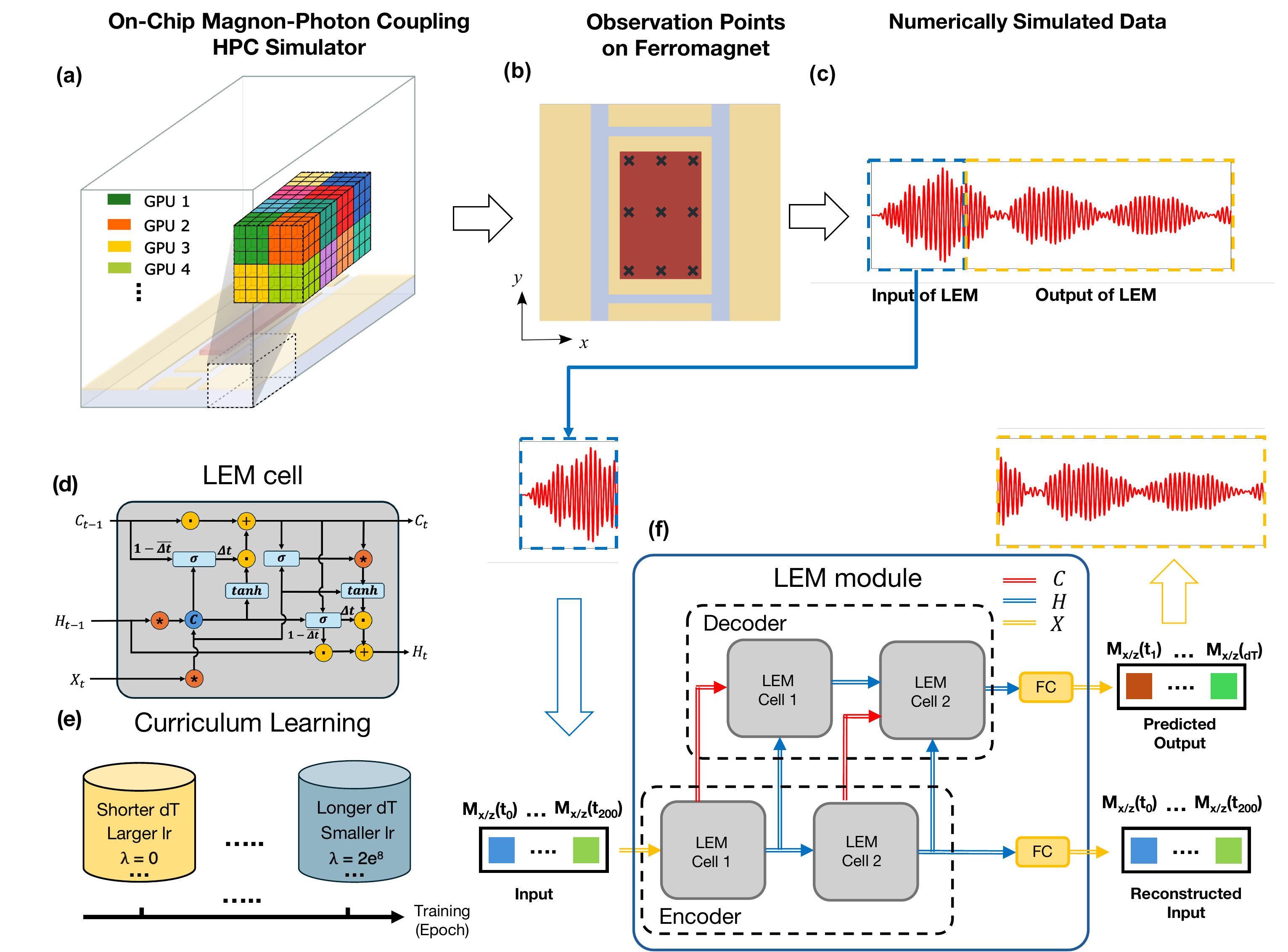}
    \caption{\textbf{The HPC and ML frameworks for magnon-photon hybrid quantum modeling}. (a) Simulation domain setup and GPU partitioning in the Maxwell–LLG coupled numerical model, ARTEMIS. The GPU parallelization leverages the grid-structured GPU library AMReX \cite{AMReX_JOSS}, achieving nearly flat weak-scaling performance.
    (b) Top view of the magnon-photon hybrid quantum circuit; the cross marks indicate probing points used for data collection. (c) Example time-series data of the magnetization \Mb~recorded at a single probing point. 
    (d) Detailed architecture of the LEM cell. The cell updates a latent state ($C_t$) and a hidden state ($H_t$) using the input $X_t$, previous states, and learnable time constants ($\Delta t$). $C_t$ captures short-term dynamics, while $H_t$ integrates transformed latent features to maintain temporal context. (e) The curriculum learning strategy used to train the ML model, with a staged schedule with progressively adjusted learning rates, batch sizes, sequence lengths, and physics-based loss weights to stabilize training and enhance convergence. 
    (f) Overall ML framework for time-series prediction, built out of the LEM-based architecture. The encoder consists of sequential LEM Cells that process the input sequence (X) into two evolving internal states: the latent state (C); and the hidden state (H). These states are used to initialize an auto-regressive decoder, which also consists of stacked LEM Cells. At each step, the decoder outputs a hidden state (H) that is passed through a shared fully connected (FC) layer to produce the predicted output.
    }
    \label{fig:fig2-HPC_ML}
\end{figure}

We employ a fully coupled dynamic Maxwell–LLG solver, ARTEMIS, based on the approach in~\cite{Yao2022}, using an explicit FDTD leap-frog scheme to update the electric field \Eb, magnetic field \Hb, and magnetization \Mb. In nonmagnetic regions, only Maxwell’s equations are solved (with \Mb \hspace{0.25em} set to zero). In magnetic regions, Maxwell’s equations and the LLG equation are solved self-consistently in a mutual feedback loop (see \nameref{subsec:methods-numerical} in \nameref{sec:methods} for details).
All field variables computed by the solver—\Eb, \Hb, and \Mb—achieve second-order accuracy, owing to the trapezoidal time-domain discretization of the LLG equation and the leap-frog time-marching scheme for Maxwell’s equations.
To handle large problem sizes, we implement both multicore and GPU parallelization for the spatial mesh, as illustrated in Figure~\ref{fig:fig2-HPC_ML}(a). We use the AMReX~\cite{AMReX_JOSS} library to partition the computational domain into non-overlapping grids, each assigned to an MPI rank. AMReX follows an MPI+X parallelization model, where \emph{X} can be OpenMP (for multicore CPUs) or CUDA (for~GPUs).

The input data for the physics-informed ML model is collected from nine discrete spatial locations, shown in Figure~\ref{fig:fig2-HPC_ML}(b). 
These points lie in the middle of the ferromagnet layer along its thickness and include the film’s center, as well as the four corners and midpoints of the four edges in the lateral plane.
More details can be found in the \nameref{subsec:methods-ml-training} section of \nameref{sec:methods}.
Figure~\ref{fig:fig2-HPC_ML}(f) illustrates the general ML framework for sequential modeling of the magnetization. 
The model is based on the Long Expressive Memory (LEM) module~\cite{rusch2022iclr}, and it comprises an LEM encoder and decoder. 
The structure of the LEM cell is illustrated in Figure~\ref{fig:fig2-HPC_ML}(d), which shows the computation flow across its internal states. 
At each time step, the input $X_t$, the previous hidden state $H_{t-1}$ and latent state $C_{t-1}$ are linearly projected and combined to generate two adaptive time-step parameters: $\Delta t$ for updating the latent state $C_t$, and $\overline{\Delta t}$ for updating the hidden state $H_t$. 
These two time-step parameters act as gating coefficients, controlling the extent to which newly computed nonlinear activations (via $\tanh$) are blended into each state. 
The latent state $C_t$ is updated first, capturing fast-evolving features from the input and hidden state. 
This updated $C_t$ is then transformed and used to update the hidden state $H_t$, which serves as the main memory for downstream prediction. 
Additional implementation details and the mathematical formulation are provided in the \nameref{subsec:methods-lem} of \nameref{sec:methods}.
The LEM module processes the input sequence (200 temporal samples) using a stacked encoder composed of two LEM cells, which evolve coupled hidden $H$ and latent states $C$ with adaptive time constants. 
The final states from the encoder are used to initialize a two-stage decoder, which begins from a zero input and autoregressively generates a fixed-length output sequence. 
At each decoding step, the internal states are updated using the same LEM dynamics, and the top-layer hidden state is passed through a fully connected layer to produce the output.
The model is trained using a composite loss function that integrates three component:
\begin{enumerate}[itemsep=0pt,parsep=0pt,topsep=0pt,partopsep=0pt]
    \item Reconstruction Loss: The Mean Squared Error (MSE) between the reconstructed input and the actual input sequence. 
    \item Prediction Loss: The MSE between the predicted and ground-truth output sequences.
    \item Physics Loss: A term enforcing physical constraints, weighted by a coefficient $\lambda$ (details provided in the \nameref{subsec:methods-loss} section of \nameref{sec:methods}).
\end{enumerate}
\noindent This combined approach enables the model to balance both data-driven objectives and physics-informed objectives.

Given the complexity of the hybrid quantum system, we incorporated Curriculum Learning \cite{2101.10382} to improve generalization and help the model better capture long-term dynamics and the underlying physics.
As shown in Figure~\ref{fig:fig2-HPC_ML}(e), this approach enables the model to handle tasks of increasing complexity over time. During training, three key hyperparameters are dynamically adjusted:
\begin{itemize}[itemsep=0pt,parsep=0pt,topsep=0pt,partopsep=0pt]
    \item[] $\Delta t$ ($dt$): The sequence length for prediction.
    \item[] $lr$: The model’s learning rate.
    \item[] $\lambda$: The weighting coefficient for the physics loss.
\end{itemize}
Training begins with simpler scenarios—shorter prediction sequences, no physics loss, and a higher learning rate—and gradually moves to more complex cases involving longer sequences, physics loss, and a lower learning rate. This staged approach ensures that the model generalizes effectively while capturing the intricate physics governing the field interactions.
Details on how the training and testing datasets were curated are provided in the \nameref{subsec:methods-data-curation} section of \nameref{sec:methods}.

\subsection{Numerically Predicted Strong Magnon-Photon Coupling Phenomena.}
\label{subsec:numerical-result}
\begin{figure}
    \centering
    \includegraphics[width=\linewidth]{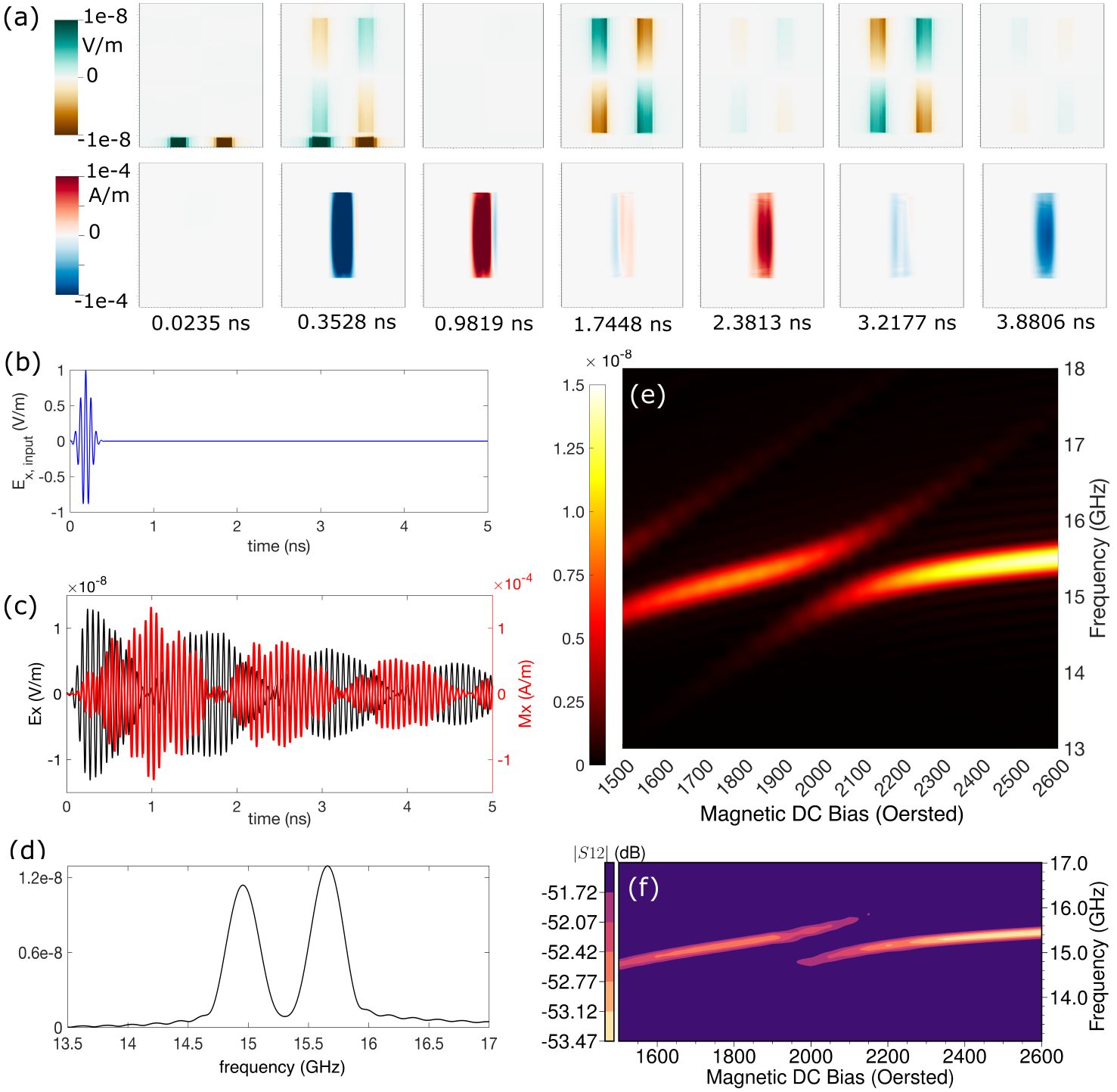}
    \caption{\textbf{Numerical modeling results} (a) Snapshots showing top-view distributions of $E_x$ (top row) and $M_x$ (bottom row) at selected time instants with an applied magnetic DC bias $H_0 = 2050$ Oersted.  Here, the $E_x$ fields are recorded at the same locations as Figure~\ref{fig:fig1-overview}(e), i.e., the vertical center plane passing through the lateral air gap of the CPW resonator, while the $M_x$ fields are recorded at the same locations as Figure~\ref{fig:fig1-overview}(f), i.e., the top surface of the ferromagnetic stripe. See Figure~\ref{fig:fig1-overview}(a) for details of the observation locations.
    Each snapshot helps illustrate how the electric field and magnetization evolve over time in their respective regions. 
    (b) Input electric field in the form of modulated Gaussian pulse. The position of the input port is depicted in Figure~\ref{fig:fig1-overview}(a).
    (c) Time sequences of $E_x$ and $M_x$ at specific spatial points. $E_x$ is recorded at the CPW resonator’s edge within its lateral air gaps (Figure~\ref{fig:fig1-overview}), while $M_x$ is recorded at the longitudinal center of the ferromagnet stripe on the top surface (Figure~\ref{fig:fig1-overview}).
    (d) Spectral $E_x$ field, obtained by Fourier-transforming the time sequence in sub-figure (c). 
    (e) Magnitude spectra of $E_x$ under varying magnetic DC bias $H_0$.
    (f) Magnitude spectra of $|S_{12}|$ parameter under the same sweeping magnetic DC bias $H_0$.}
    \label{fig:fig3-num-result}
\end{figure}

Our Maxwell-LLG dynamical modeling has predicted the dynamically evolving and interacting electric field and magnetization, shown in Figure~\ref{fig:fig3-num-result}. 
Figure~\ref{fig:fig3-num-result}(a) presents top-view snapshots of $E_x$ (top row) and $M_x$ (bottom row) at selected time instants, with an applied magnetic DC bias $H_0 = 2050 $ Oersted.
According to the Kittel's law, under such $H_0$, the FMR frequency is $f_{\textrm{m}} = \omega_\textrm{m}/2\pi = (\gamma / 2 \pi) \sqrt{H_0(H_0+4\pi M_S)} = 15.027 $ GHz, detuned from the photon frequency $f_\textrm{p}$ by about $\Delta = 200$ MHz. 
This small detune enables strong mode hybridization and, consequently, strong magnon-photon coupling.
The $E_x$ fields are observed at the same location described in Figure~\ref{fig:fig1-overview}(e), i.e., along the vertical center plane passing through the CPW resonator’s lateral air gap. 
The $M_x$ fields are observed at the same location described in Figure~\ref{fig:fig1-overview}(f), i.e., on the top surface of the ferromagnetic stripe.
For additional details on these observation points, see Figure~\ref{fig:fig1-overview}(a).
At 0.0235 ns, the electric field appears only in the feeding patches, without exciting resonant modes (photons) in the CPW resonator, and no dynamic magnetization is observed in the ferromagnet.
By 0.3528 ns, both $E_x$ and $M_x$ emerge in the resonator.
Starting at 0.9819 ns, even though the input field excitation has vanished (the input signal completely  decays by 0.5 ns, as shown in Figure~\ref{fig:fig3-num-result}(b)), oscillations of $E_x$ and $M_x$ persist within the CPW resonator. 
Notably, $E_x$ peaks near the resonator’s ends, whereas $M_x$ peaks at the center.
Both fields exhibit well-defined oscillations with only minor decay, indicating low damping—a result of minimal Gilbert damping, small metallic losses, and limited electromagnetic leakage into air or substrate.
Furthermore, $E_x$ and $M_x$ alternate in reaching their standing-wave maxima: when $E_x$ is at a peak, $M_x$ is at a node, and vice versa. 
This phase alternation underscores strong energy exchange between the magnon and photon polaritons, more clearly seen in the transient response of Figure~\ref{fig:fig3-num-result}(c), and the yielded mode splitting in the Fourier transformed $E_x$ spectra shown in Figure~\ref{fig:fig3-num-result}(d). 
Figure~\ref{fig:fig3-num-result}(e) visualizes the spectra of $E_x$ magnitude under various $H_0$, revealing a clear anti-crossing behavior, indicating coherent coupling between magnons and photons~\cite{ZareRameshti2022}. 
The simulated spectrum matches theoretical analysis as shown in Extended Data Figure~\ref{fig:ext1-theory}.
The $S$-parameter ($|S_{12}|$) spectra shown in Figure~\ref{fig:fig3-num-result}(f) is converted from the $E_x$ field following the workflow based on field calculation in ~\cite{Sawant2023}, indicating the transmission rate from one end to the other of the microwave resonator, also showing an anti-crossing spectra.  
Both figures show mode splitting $g_\textrm{mp} = 670$ MHz, where $g_\textrm{mp}$ is the total magnon-photon coupling strength, defined as $g_\textrm{mp} = g_0 \sqrt{N}$, with $g_0$ defined as the coupling strength between photons and individual spins in the ferromagnet, and $N$ is the number of spins in the ferromagnet~\cite{MagnonPhotonTheory, Zhang2014}. 
$g_0$ contributes to the total Hamiltonian of the system:
\begin{equation}
\hat{H}
= \hbar \omega_\textrm{p}\Bigl(\hat{a}_{p}^{\dagger}\hat{a}_{p} + \tfrac12\Bigr)
- \omega_\textrm{m}(H_{0})\,\hat{S}_{y}
+ g_{0}\bigl(\hat{S}_{+}\,\hat{a}_{p}^{\dagger} \;+\; \hat{S}_{-}\,\hat{a}_{p}\bigr),
\label{eq:hamiltonian}
\end{equation}
\noindent where $\hat{a}_{p}^{\dagger}  (\hat{a}_{p})$ is the creation (annihilation) operator of microwave photon modes in the CPW resonator, and $\hat{\textbf{S}} = -\hat{x}\frac{1}{2}(\hat{S}_{+}+\hat{S}_{-}) + \hat{y}\hat{S}_{y} + \hat{z}\frac{1}{2}(\hat{S}_{+}-\hat{S}_{-})$ is the macrospin operator, with $\hat{S}_{+}$($\hat{S}_{-}$) raising (lowering) the $y$ component of the
macrospin.
The eigen frequencies of hybrid system can be measured from Figure~\ref{fig:fig3-num-result}(e) and (f) as 14.95 GHz and 15.65 GHz, matching the theoretical prediction of ~\cite{Huebl2013}:
\begin{equation}
\omega_{\pm}
= \omega_{p} + \frac{\Delta}{2}
\;\pm\;
\frac{\sqrt{\Delta^{2} + 4\,g_\textrm{mp}^{2}}}{2} .
\label{eq:omega_pm}
\end{equation}

\noindent 
One unique advantage of numerical simulation is the ability to generate signals on demand, including those with intensities that are experimentally unattainable. 
Extended Data Figure~\ref{fig:ext2-nonlinear} displays the electric-field spectra as the excitation intensity is increased. It demonstrates that, in the absence of significant spin-spin interactions (e.g., exchange coupling), progressively stronger microwave magnetic fields suppress the magnon dynamics and leave only the cavity photon mode.  Such behaviour arises because, at high powers, nonlinear spin-wave interactions—particularly three-magnon splitting—induce a Suhl instability that transfers energy from the uniform ferromagnetic resonance to non-uniform spin waves; this saturation of the Kittel mode causes the magnon–photon anticrossing gap to close and yields a single photon-like resonance, a phenomenon sometimes described as the high-power cavity magnon-polaritons regime~\cite{nonlinear2023}. 
For more theoretical analysis and validation under both linear and nonlinear regimes, refer to the \textbf{Supplementary Materials}. 
These results collectively demonstrate the model's capability in predicting the intricate hybrid quantum coupling, especially in complex, realistic integrated chip setup. 

\subsection{ML Surrogate Accuracy and Generalization.}
\label{subsec:ml-result}
\begin{figure}
    \centering
    \includegraphics[width=\linewidth]{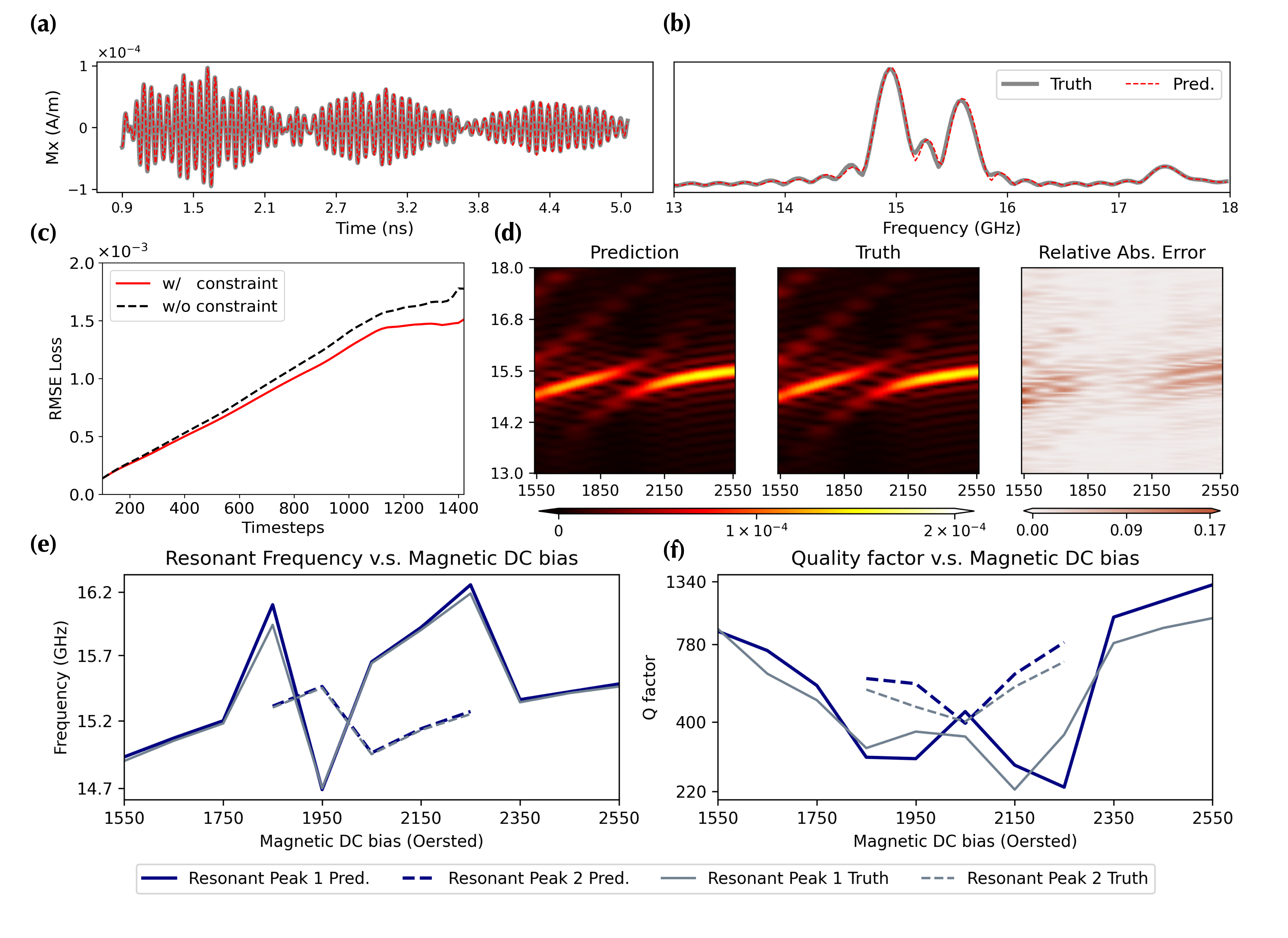}
    \caption{\textbf{Physics-informed ML time sequence prediction at one single spatial probing point.} 
    (a) Time-series of the probed $M_x$ field: ML prediction versus numerical ground truth. 
    (b) Fourier spectra of the same traces as (a). 
    (c) The root mean squared error (RMSE) as a function of forcast length, with and without the physical loss. 
    (d) Magnitude spectra of $M_x$ under varying magnetic DC bias $H_0$. The x-axis is the bias magnetic field in Oersted, and the y-axis is the frequency in GHz; color indicates field amplitude. 
    (e) Hybrid-mode eigen frequecies extracted with ESPRIT~\cite{esprit}-calculated eigen frequencies of the hybrid modes.  
    (f) Corresponding ESPRIT-calculated $Q$ factors of the hybrid modes.}
    \label{fig:fig4-ML_single}
\end{figure}

\begin{figure}
    \centering
    \includegraphics[width=\linewidth]{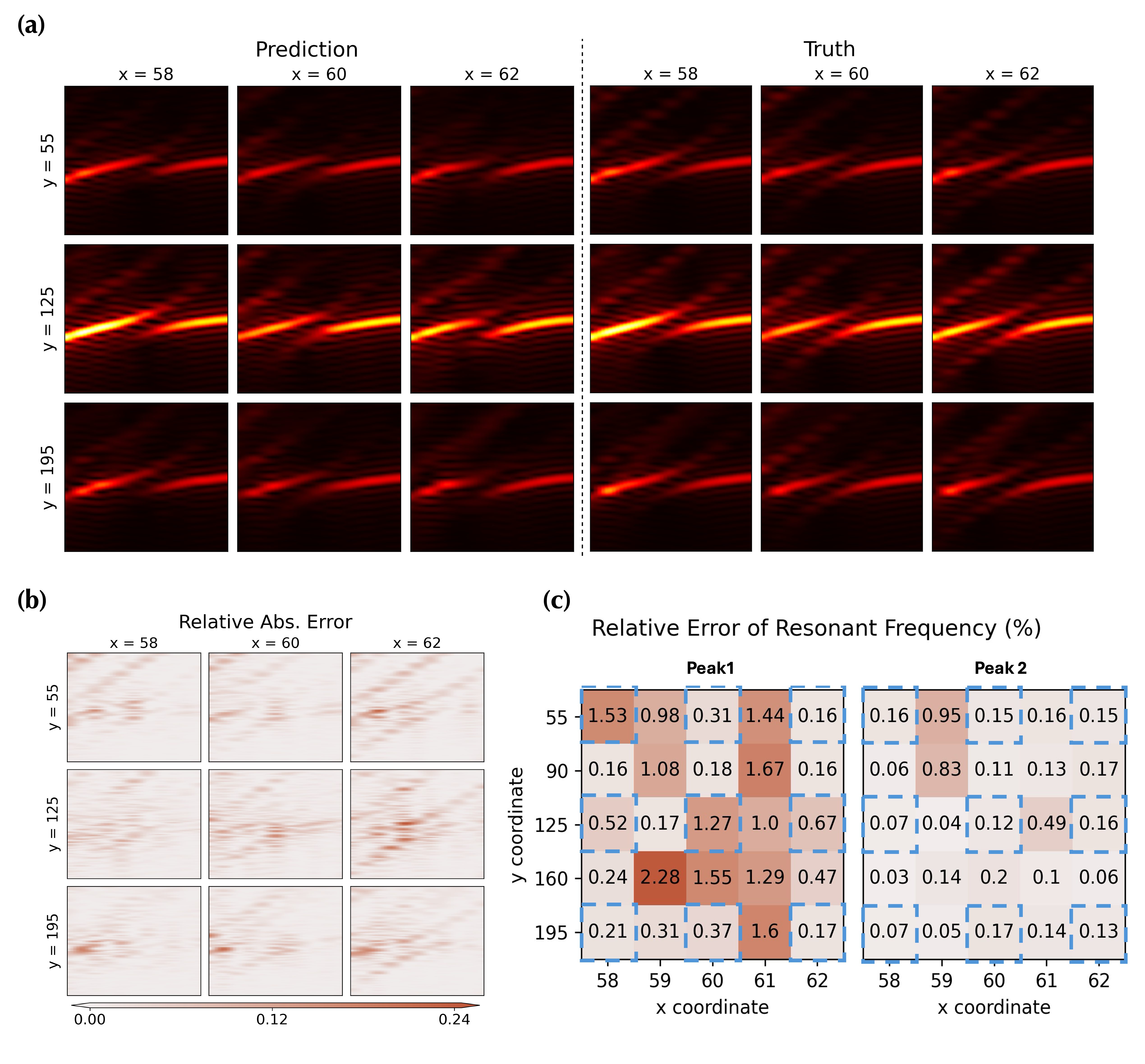}
    \caption{\textbf{Physics-informed ML predictions at multiple probe points.} 
    (a) Magnitude spectra of the $M_x$ field at the nine in-distribution (ID) locations: model output (left) and numerical reference (right).
    (b) Relative-error map between the spectra in (a).
    (c) Relative error of the two resonance frequencies extracted at each of the 25 probe points; blue dotted markers denote the nine ID points, the remaining points are out-of-distribution (OOD).}
    \label{fig:ML_res2}
\end{figure}

We evaluate the performance of our proposed ML model for modeling long-horizon dynamical fields under two scenarios: predictions at a single spatial location (Figure~\ref{fig:fig4-ML_single}); and multiple probing locations (Figure~\ref{fig:ML_res2}). 
The model is trained using limited input data and physics constraint represented by a specific physical loss term (refer to \nameref{subsec:methods-loss} in \nameref{sec:methods} for details). 
A representative example is shown in Figure~\ref{fig:fig4-ML_single}(a), where the ML model is given only 200 temporal samples (0.6 ns) as input and predicts the long subsequent 1480 temporal samples (4.1 ns) under a DC magnetic bias of 2050 Oersted. 
The frequency spectrum in Figure \ref{fig:fig4-ML_single}(b) and Figure \ref{fig:fig4-ML_single}(d) shows that the ML surrogate reproduces the hybridized polaritons' characteristics, confirming it has learned the underlying physics rather than simply fitting short‑term temporal trends.
An ablation test (Fig. \ref{fig:fig4-ML_single}(c)) shows physics‑informed training achieves lower long‑horizon error than purely data‑driven models, while requiring fewer samples and training steps.
Using the ESPRIT algorithm~\cite{esprit}, we extracted the two polariton frequencies and their $Q$‑factors directly from the time traces in Figure \ref{fig:fig4-ML_single}(a); the results are plotted in panels (e) and (f).
Frequencies match almost exactly, while small $Q$‑factor offsets persist—typical for high‑$Q$ systems where tiny signal errors and hyper‑parameter choices strongly affect the fitted decay rate.
Even so, the physics‑informed model reproduces both resonant modes with high fidelity, confirming its accuracy for precision‑sensitive metrics.

To further test spatial generalization, we move from a single‑spatial-probing model to a multi‑spatial-probing network with a larger hidden layer. 
Training uses data from nine monitored points; held‑out temporal samples from these same spatial points serve as in‑distribution (ID) tests, while signals from 16 previously unseen spatial points provide out‑of‑distribution (OOD) tests.
The nine color maps on the left ``Prediction'' panel of Figure~\ref{fig:ML_res2}(a) show the $M_x$ spectra predicted at ID probing points, and they are visually indistinguishable from the numerical reference on the right ``Truth'' panel. 
The corresponding relative-error map in Figure~\ref{fig:ML_res2}(b) confirms that discrepancies stay negligible, particularly in the polariton resonance bands. 
We also applied the network to 16 probing points that were withheld from training (OOD), and generated similar spectra as the ID data. 
Figure~\ref{fig:ML_res2}(c) plots the mean relative error for the first (“Peak 1”) and second (“Peak 2”) polariton modes; blue dotted lines mark the nine ID positions. 
Across all 25 probe points the physics‑informed network keeps the resonance‑frequency error below $2.3\%$, and the OOD sites perform on par with the ID ones. 
The uniformly low errors confirm true spatial generalization, enabled by the physics‑based loss terms. 
Consequently, the model predicts field patterns and resonance spectra at unseen locations without any additional retraining.

\section{Discussion}
\label{sec:discussion}
\subsection{GPU Scalability.}
\label{subsec:scaling}

\begin{figure}
    \centering
    \includegraphics[width=\linewidth]{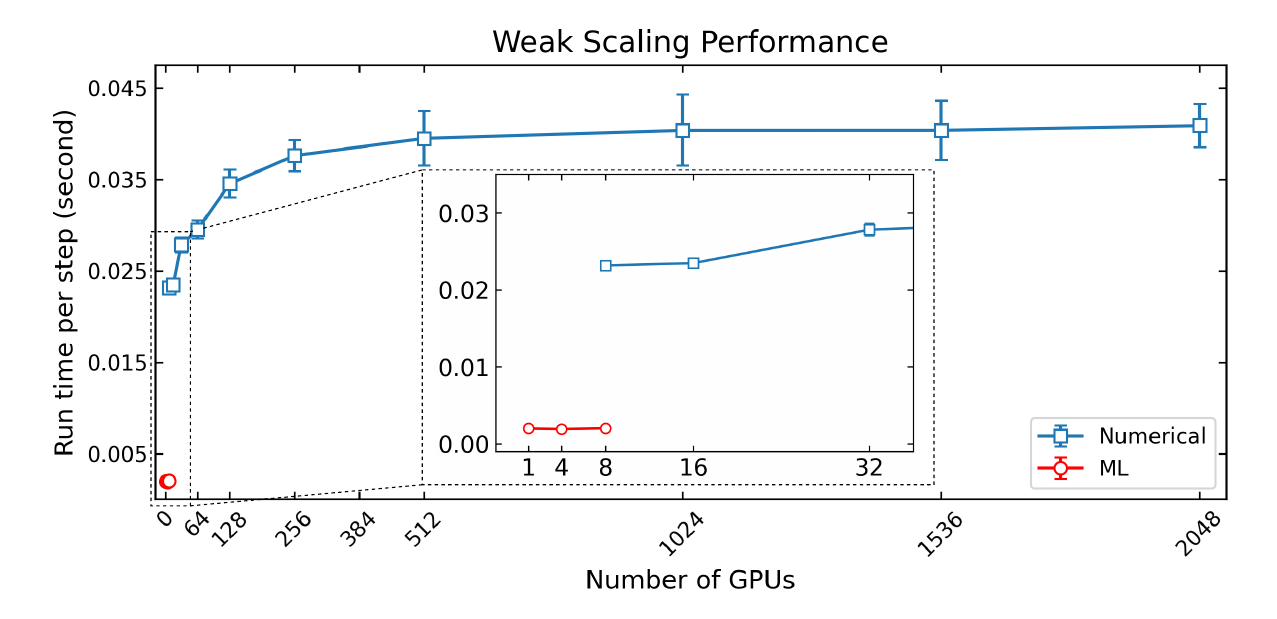}
    \caption{Weak scaling for simulations on the NERSC Perlmutter GPU (MPI+CUDA) system. Note that one Perlmutter node contains 4 GPUs. The weak scaling is nearly perfect for the Perlmutter simulations past the 16-node threshold up to 2048 GPUs (512 nodes). Details of numerical accuracy and performance tests can be found in the \nameref{subsec:methods-scaling} section in \nameref{sec:methods}. }
    \label{fig:fig5-scaling}
\end{figure}

Figure~\ref{fig:fig5-scaling} shows the weak-scaling results for our ARTEMIS multi-core/GPU solver, where each GPU advances a fixed mesh size. 
Performance is essentially ideal from 512 to 2048 GPUs, far beyond the scaling range of existing commercial or research codes that tackle Maxwell-LLG coupling, making full-chip modeling of hybrid quantum circuits and interconnects practical. 
For comparison, the same figure includes the cost of the physics-informed ML surrogate when inferring a single probe point on 1, 4, and 8 GPUs. 
The run time is normalized as $t_\textrm{inf} = T N_p/(N_t B)$, where $T$ is the total wall-clock time, $N_t$ the number of predicted steps, $B$ the batch size, and $N_p$ the number of spatial points, allowing direct comparison with the per-step cost of the numerical solver. 
Because the surrogate operates directly on time-series data and needs no full-3D mesh, its per-step cost is orders of magnitude lower --- an advantage when only a handful of spatial points are of interest, as in most experimental measurements. 

\subsection{Future Directions.}
\label{subsec:outlook}
We demonstrated hybrid quantum simulation using a magnon–photon testbed, yet the framework is readily extensible to other hybrid magnonic, cavity electrodynamic, and phononic systems ~\cite{Chen2018,Bossini2021,Hwang2024,Aspelmeyer2014}.
Ongoing work therefore targets a generalized engine for quantum communication, storage, and computation that natively supports these additional interactions. 
Our current ML surrogate treats each mesh point as an independent one-dimensional time series, limiting its view of the device’s collective field behavior. 
We therefore aim to develop spatiotemporal models that learn the coupled evolution of two- and three-dimensional fields, preserve spatial coherence, and predict the full device response with high fidelity.

\section{Methods}\label{sec:methods}

\subsection{Numerical Model.}
\label{subsec:methods-numerical}

In the entire simulation domain, we solve the full-form of the dynamic Maxwell's equations, and Ampere and Faraday's laws, given as:
\begin{equation}
\nabla\times\Hb = \epsilon_0\epsilon_r\frac{\partial\Eb}{\partial t} + \sigma \Eb,\label{eq:Ampere}
\end{equation}
\begin{equation}
\nabla\times\Eb = -\frac{\partial\Bb}{\partial t} ,
\label{eq:Faraday}
\end{equation}
where $\Eb$ is the electric field, $\Bb$ the magnetic flux density, $\Hb$ the magnetic-field intensity, and $\Jb_{\rm src}$ an impressed current source. 
Material parameters are the conductivity $\sigma$, vacuum permittivity $\epsilon_0$, relative permittivity $\epsilon_r$, and vacuum permeability $\mu_0$. 
The constitutive law linking $\Hb$ and $\Bb$ is 
$\Bb = \mu\Hb = \mu_0(\Hb + \Mb),$
with $\Mb$ the magnetization density and $\mu=\mu_0\mu_r$ the permeability.  
This makes Equation~\ref{eq:Faraday} into $\nabla\times\Eb = -{\partial\Hb}/{\partial t} - {\partial\Mb}/{\partial t}$.
For non-magnetic media, $\mu_r\!\approx\!1$ since $\Mb\!\approx\!0$.  
In ferromagnets, $\Mb$ must be evolved (e.g., via the Landau–Lifshitz–Gilbert (LLG) equation) to capture magnon dynamics. 
The LLG equation is expressed as:
\begin{equation}
\frac{\partial\Mb}{\partial t} = \mu_0 \gamma (\Mb \times \Heff) + \frac{\alpha}{|\Mb|} (\Mb \times \frac{\partial\Mb}{\partial t}),
\label{eq:LLG}
\end{equation}
where $\gamma = -1.759 \times 10^{11}$ C/kg is the gyromagnetic ratio and $\alpha$ is the dimensionless Gilbert damping constant ($10^{-5}$ to $10^{-3}$ in typical ferromagnets).
The effective field $\Heff$ collects all magnetic interactions. 
Here we retain two terms essential for magnon-photon coupling: the time-dependent EM field $\Hb_\mathrm{EM}$, and the static bias $\Hb_0$.
Because the ferromagnet is driven close to saturation, the magnetization magnitude is taken constant, $|\Mb| = M_s$, where $M_s$ is the (spatially dependent) saturation magnetization.
We employ an iterative scheme that uses a trapezoidal discretization of the vector form of the LLG equation (\ref{eq:LLG}) described in \cite{Yao2022}, to solve for $\Mb^{n+1,r}$, where $n$ denotes the time step and $r$ is the counter index for the iterative procedure:
\begin{equation} \label{eq:llg_tra4}
\Mb^{n+1,r} = 
\frac{\boldsymbol{b} + (\boldsymbol{a} \cdot \boldsymbol{b}) \boldsymbol{a} - \boldsymbol{a} \times \boldsymbol{b}}{1+|\boldsymbol{a}|^{2}},
\end{equation}
where
\begin{equation} \label{eq:llg_tra5}
\boldsymbol{a} = - \left[ \frac{ \mu_0 |\gamma| \Delta t}{4} \Hb^{n+1,r-1}_{\rm eff}
+ \frac{\alpha}{M_s}\Mb^n
\right],
\end{equation}
\begin{equation} \label{eq:llg_tra6}
\boldsymbol{b} = \Mb^{n} - \frac{\mu_0 |\gamma| \Delta t}{4} (\Mb^{n} \times \Hb^{n}_{\rm eff}). 
\end{equation}
We normalize $\Mb^{n+1,r} = \mathcal{N}(\Mb^{n+1,r})$ and then compute $\Hb^{n+1,r}$ by integrating Faraday's law following the standard FDTD procedure, with an updated approximation for the time-derivative of $\Mb$,
\begin{equation}
\Hb^{n+1,r} = \Hb^n + \Mb^n - \Mb^{n+1,r} - \frac{\Delta t}{2\mu_0}\nabla\times\Eb^n.\label{eq:iterate_H}
\end{equation}
We continue to iterate over $r$ using equations (\ref{eq:llg_tra4}) and (\ref{eq:iterate_H}), until the maximum value of $(\Mb^{n+1,r} - \Mb^{n+1,r-1})/M_s$ over the entire domain is smaller than a user-defined tolerance that we choose to be $10^{-6}$.
After the iterations converge, we set $(\Hb,\Mb)^{n+1} = (\Hb,\Mb)^{n+1,r}$ and proceed to computing $\Eb^{n+1}$ by discretizing Ampere's law (\ref{eq:Ampere}) using time-centered representations of the magnetic field and current,
\begin{equation}
\Eb^{n+1} = \left(\frac{\sigma}{2} + \frac{\epsilon}{\Delta t}\right)^{-1}\left[\nabla\times\Hb^{n+1} - \left(\frac{\sigma}{2} - \frac{\epsilon}{\Delta t}\right)\Eb^n - \Jb_{\rm src}^{n+1}\right].
\end{equation}
Note that in regions where $M_s=0$, we set $\Mb=0$, skip this iterative scheme, and instead integrate $\Hb$ using equation (\ref{eq:iterate_H}), with ~$\mu_0$.

\subsection{Numerical Solver Implementation.}
\label{subsec:methods-amrex}
We build our solver on AMReX~\cite{AMReX_JOSS}, the DOE Exascale Computing Project’s block-structured adaptive mesh refinement (AMR) library. 
AMReX provides structured-grid PDE infrastructure with performance portability: the domain is decomposed into non-overlapping boxes (``grids'') that are distributed across MPI ranks, and execution follows an MPI+X model: X=OpenMP on multicore CPUs; or X=CUDA on NVIDIA GPUs. 
A single front-end code path uses AMReX loop abstractions, which select the appropriate back end at compile time.
Each rank updates only the data it owns using triply nested cell loops. 
In the pure-MPI case, these are standard i-j-k loops; with MPI+OpenMP the loops are further tiled and assigned to threads; with MPI+CUDA AMReX launches device kernels that map cells to CUDA threads. 
AMReX manages data residency so arrays remain on the GPU whenever possible, minimizing host–device traffic; inter-rank communication and host/GPU transfers are therefore limited to ghost-cell exchanges performed a small number of times per time step.
The efficiency and scalability of our code using MPI+CUDA simulations on NERSC Perlmutter systems is shown in Figure~\ref{fig:fig5-scaling}.
Data from the simulation can be efficiently written using a user-defined number of MPI ranks, to prevent overwhelming the system with simultaneous writes. 
Visualization was performed with the publicly-available software package, VisIt \cite{VisIt}.


\subsection{Datasets Curation.}
\label{subsec:methods-data-curation}
The dataset for our ML model was generated using the ARTEMIS simulator introduced in sections \nameref{subsec:methods-amrex} and \nameref{subsec:methods-numerical}. 
We monitored the $x$ and $z$ component of magnetization ($M_x$ and $M_z$) at discrete spatial points across a defined domain, creating a dataset with sparse spatial sampling.
The $x$-indices range from $58 $ to $62 $ with index interval of 1, corresponding to the x location of 290 $\mu m$ to 310 $\mu m$.
The $y$-indices spanned $55$ to $195$ with a sampling interval of 35, corresponding to 1045 $\mu m$ and 3705 $\mu m$. 
The $z$-coordinate was fixed at 505 $\mu m$ (and thus the $z$-index is also 505), corresponding to the center of the NiFe magnetic strip along the $z$-axis, resulting in a total of 25 spatial sampling points.
Simulations were conducted under 20 distinct external DC magnetic bias conditions, ranging from 1550 Oersted to 2550 Oersted, sampled uniformly at intervals of 50 Oersted.
For the single-location scenario, we used the temporal field data $M_x$ and $M_z$ from a specific spatial location $(60,125,505)$, located at the center of the NiFe magnetic material strip. Four trajectories corresponding to external magnetic biases of 2050, 2250, 2350, and 2550 Oersted were reserved for testing, while the remaining trajectories were used for training.
For multiple-location secario, the spatial locations include 9 monitoring points: \([x, y, 505]\), where \(x \in \{58, 60, 62\}\) \(\mu\)m and \(y \in \{55, 125, 195\}\) \(\mu\)m. For each spatial point, the training trajectories are chosen from the magnetic field \(H_0\) across the range \([1500, 1600, \dots, 2600]\) Oersted, using a step size of 100.
Each trajectory is preprocessed by truncating the first 160 steps, which is the ramping-up stage of the input signal, and downsampling by a factor of 1000. 
The data is normalized using min-max normalization. 


\subsection{LEM Cell.}
\label{subsec:methods-lem}
We use the Long Expressive Memory (LEM) model~\cite{rusch2022iclr} as a robust approach for modeling complex temporal dynamics. LEM is specifically designed to capture multi-scale and long-term behaviors in systems characterized by intricate temporal structures. Its architecture offers a mathematical foundation for handling sequences with varying temporal scales, while ensuring gradient stability and high expressivity. EM fields, which exhibit a combination of rapid oscillations (e.g., high-frequency wave propagation) and slower dynamics (e.g., energy dissipation), present a challenging time-series modeling task. The LEM framework is well-suited for our scientific task due to the capability to effectively learn and predict diverse and interdependent behaviors within EM dynamics.
The LEM model is rooted in multi-scale Ordinary Differential Equations (ODEs) that can be used to capture fast and slow dynamics. Let us consider a general formulation of a two-scale ODE system \cite{rusch2022iclr}, given as
\begin{equation}
\begin{split}
\label{eq:lem}
    \frac{d \mathbf{y}}{dt} &= \tau_y \left( \sigma(\mathbf{W}_y \mathbf{z} + \mathbf{V}_y \mathbf{u} + \mathbf{b}_y) - \mathbf{y} \right), \\
    \frac{d \mathbf{z}}{dt} &= \tau_z \left( \sigma(\mathbf{W}_z \mathbf{y} + \mathbf{V}_z \mathbf{u} + \mathbf{b}_z) - \mathbf{z} \right).
\end{split}
\end{equation}
Here, $\mathbf{y}(t) \in \mathbb{R}^d$  and  $\mathbf{z}(t) \in \mathbb{R}^d$ represent the slow and fast variables, $\mathbf{u}(t) \in \mathbb{R}^m$  is the input signal, and $\{\tau_y, \tau_z\}$ are time-scale parameters ($0 < \tau_y \leq \tau_z \leq 1$). We define $\{\mathbf{W}_y, \mathbf{W}_z, \mathbf{V}_y, \mathbf{V}_z\}$ as weight matrices and $\{\mathbf{b}_y, \mathbf{b}_z\}$ as bias vectors. $\sigma(\cdot)$ denotes the nonlinear activation function (e.g., hyperbolic tangent function). This formulation can be further generalized to represent multiple temporal scales by incorporating additional gating functions. These gating mechanisms, based on adaptive weights $\{\mathbf{W}_1, \mathbf{W}_2, \mathbf{V}_1, \mathbf{V}_2\}$ and biases $\{\mathbf{b}_1, \mathbf{b}_2\}$, allow the model to dynamically learn and modulate across various time scales. Therefore, the LEM model is capable of capturing more intricate dependencies and temporal patterns. To ensure practical applicability, LEM discretizes the continuous multi-scale ODE system using an implicit-explicit (IMEX) time-stepping scheme, which can be formulated as
\begin{equation}
    \begin{split}
    \label{eq:discrete_lem}
    \Delta \mathbf{t}_n &= \Delta t \, \hat{\sigma}(\mathbf{W}_1 \mathbf{y}_{n-1} + \mathbf{V}_1 \mathbf{u}_n + \mathbf{b}_1), \\
    \overline{\Delta \mathbf{t}}_n &= \Delta t \, \hat{\sigma}(\mathbf{W}_2 \mathbf{y}_{n-1} + \mathbf{V}_2 \mathbf{u}_n + \mathbf{b}_2), \\
    \mathbf{z}_n &= (1 - \Delta \mathbf{t}_n) \odot \mathbf{z}_{n-1} + \Delta \mathbf{t}_n \odot \sigma(\mathbf{W}_z \mathbf{y}_{n-1} + \mathbf{V}_z \mathbf{u}_n + \mathbf{b}_z), \\
    \mathbf{y}_n &= (1 - \overline{\Delta \mathbf{t}}_n) \odot \mathbf{y}_{n-1} + \overline{\Delta \mathbf{t}}_n \odot \sigma(\mathbf{W}_y \mathbf{z}_n + \mathbf{V}_y \mathbf{u}_n + \mathbf{b}_y).
    \end{split}
\end{equation}
Here, $\odot$ denotes element-wise multiplication, $\hat{\sigma}(\cdot)$ is the sigmoid activation function, $\Delta t > 0$ is a tunable parameter, and $n$ represents the discrete time step within $[1,N]$. $\{\mathbf{y}_n, \mathbf{z}_n\} \in \mathbb{R}^d$ and $\mathbf{u}_n \in \mathbb{R}^m$ denote two hidden states and input variable, respectively. The final output state $\mathbf{o}_n \in \mathbb{R}^o$ is augmented with a linear operation $\mathcal{W}_y \in \mathbb{R}^{o\times d}$, where $\mathbf{o}_n=\mathcal{W}_y \mathbf{y}_n$. 


\subsection{ML Training Implementation.}
\label{subsec:methods-ml-training}

We implement curriculum training to enhance model performance. This approach is widely adopted in scientific ML since it allows the ML model to gradually adapt to increasing levels of complexity and facilitates the effective learning of long-term temporal dynamics.
For the single probing point training, the model is trained for a total of 1250 epochs with the hidden unit number of the LEM cell set to 64. Throughout the training, the learning rate is gradually decreased from 0.001 to $1.25\times10^{-4}$, the predicted sequence length is adjusted from 50 to 400, the batch size increases from 64 to 512, and the regularization parameter for the physics loss is progressively raised from 0 to $2.5\times 10^6$. Additionally, the training sequence size alternates between 100 and 200 at different stages to optimize learning dynamics. The training protocol is generally set the same for the multiple probing point training, except that the predicted sequence length is adjusted from 50 to 500.


\subsection{Physics Loss.} 
\label{subsec:methods-loss}
Physics-informed ML has been shown to effectively preserve physical properties in various domains~\cite{karniadakis2021physics}. 
Physics-informed loss function is adopted into the LEM framework to further improve the model's ability to learn the underlying physics. 
We adopted the small-signal approximation (the dynamic magnetization in the bias direction is negligible, i.e. $M_y \approx 0$) in the LLG equation when deriving the physics loss term, which is 
\begin{equation}\label{eq:ml_tra1}
    \mathcal{L}_{\text{phy}} = \|\text{Residual}_{M_x}\|^2 + \|\text{Residual}_{M_z}\|^2,
\end{equation}
where 
\begin{equation}\label{eq:ml_tra2}
    \text{Residual}_{M_x} = M^{n+1}_x(\textrm{LLG}) - {M^{n+1}_x}(\textrm{ML}),
\end{equation}
\begin{equation}\label{eq:ml_tra3}
    \text{Residual}_{M_z} = M^{n+1}_z(\textrm{LLG}) - {M^{n+1}_z}(\textrm{ML}).
\end{equation}
The analytical derivatives \(M_x(\textrm{LLG})\) and \(M_z(\textrm{LLG})\) are the \(x\)- and \(z\)-components derived from the trapezoidal discretization of the LLG equation, as given in Eqs.~(\ref{eq:llg_tra4})–(\ref{eq:llg_tra6}).
Given a sequence of network predictions for \(n=0,\dots,T\), these residuals are evaluated for \(n=0,\dots,T{-}1\).

\vspace{4pt}
From the ML side, we estimate the time derivatives from the predicted sequence using a finite forward–Euler discretization:
\begin{align}
\frac{\partial M_x^n}{\partial t}(\mathrm{ML}) &= \frac{M_x^{n+1} - M_x^{n}}{\Delta t_\mathrm{ML}}, \\
\frac{\partial M_z^n}{\partial t}(\mathrm{ML}) &= \frac{M_z^{n+1} - M_z^{n}}{\Delta t_\mathrm{ML}}.
\end{align}
Given these derivatives, the ML one-step \emph{next values} are
\begin{equation}
M_x^{n+1}(\mathrm{ML}) = M_x^n + \Delta t_\mathrm{ML}\,\frac{\partial M_x^n}{\partial t}, \qquad
M_z^{n+1}(\mathrm{ML}) = M_z^n + \Delta t_\mathrm{ML}\,\frac{\partial M_z^n}{\partial t}.
\label{eq:ml_next}
\end{equation}
Here \(\Delta t_\mathrm{ML}\) denotes the sampling interval of the training data, which may differ from the \(\Delta t\) utilized in the numerical solver, the magnetization values \(M_{x,z}^n\) and \(M_{x,z}^{n+1}\) 
are the neural network predictions at two consecutive timesteps.

The total training objective is
\[
\mathcal{L}_\mathrm{total} = \mathcal{L}_\mathrm{data} + \lambda\,\mathcal{L}_\mathrm{phy}.
\]
The loss term $\mathcal{L}_{\text{phy}}$ penalizes deviations from the LLG dynamics, ensuring that the predicted magnetization components \(M_x\) and \(M_z\) closely follow the physical laws governing their evolution. By minimizing this loss during training, the model learns to approximate the time evolution of magnetization accurately, while respecting the underlying physics.

%


\subsection{Code Scaling Performance.}
\label{subsec:methods-scaling}
We assess the performance and scalability of our code on Perlmutter, the NERSC flagship supercomputer. 
Perlmutter comprises 1,792 GPU-accelerated nodes, each with 4 NVIDIA A100 (40 or 80 GB) GPUs. 
We perform weak-scaling tests using an MPI+CUDA strategy with one MPI rank per GPU.
To ensure proper load balancing, the weak-scaling cases use a continuous CPW line and ferromagnetic strip along $y$, so no cavity resonances form; instead, propagating waves traverse the CPW. 
This yields uniform compute and memory load per GPU. 
Our baseline domain is $N_x\times N_y\times N_z = 120\times 250\times 250$ (7.5M cells) on 8 GPUs. For weak scaling, we increase $N_y$ proportionally to the GPU count to keep the cells per GPU fixed; e.g., 32 GPUs for $120\times 1000\times 250$ (30M cells), matching the main-manuscript problem size.
Weak-scaling results (average simulation time per time step vs. number of nodes ) are shown in Figure~\ref{fig:fig5-scaling}. 
Initialization time is excluded (one-time cost). 
We tested up to 512 nodes (2,048 GPUs) and observe near-ideal scaling from 128 to 512 nodes. 
This behavior reflects that with a fixed per-GPU subdomain, both computation and halo-exchange volumes per GPU remain constant; thus inter-node communication overhead stays small relative to compute for the problem sizes considered.
Details of the machine-learning surrogate scaling are provided in Section~\ref{subsec:scaling}.


\section{Data availability}
\label{sec:data-avail}
The datasets generated during and/or analysed during the current study will be available in the zenodo repository 
upon publication.

\section{Code availability}
\label{sec:code-avail}
All codes are publicly available on the open-source github platform. The AMReX library can be obtained at \href{https://github.com/AMReX-Codes/amrex}{AMReX}, hash {\tt 2ee8ae270} and this code can be obtained at
\href{https://github.com/AMReX-Microelectronics/Artemis}{ARTEMIS}, hash {\tt 151b5a32f}.
The ML code will be available online upon publication.

\section{Acknowledgments}
\label{sec:code-ack}
This material is based upon work supported by the U.S. Department of Energy, Office of Science, National Quantum Information Science Research Centers, Quantum Systems Accelerator (Award No. DE-SCL0000121).
Additional support is acknowledged from the U.S. Department of Energy, Office of Science, Advanced Scientific Computing Research Program, Microelectronics Science Research Center, under contract No. DE-AC02-05-CH11231.
This work was also supported in part by previous breakthroughs obtained through the Laboratory Directed Research and Development Program of Lawrence Berkeley National Laboratory under U.S. Department of Energy Contract No. DE-AC02-05CH11231.
The work at the University of Wisconsin-Madison was supported by the U.S. Department of Energy, Office of Science, Basic Energy Sciences, under Award Number DE-SC0020145 as part of the Computational Materials Sciences Program (Y.Z. and J.-M.H.).
This research used resources of the National Energy Research Scientific Computing Center (NERSC), a U.S. Department of Energy Office of Science User Facility operated under Contract No. DE-AC02-05CH11231. 

\section{Author contributions}
\label{sec:authors}
Z.Y. conceived the study and supervised the research.
Z.Y. and Y.T. led the manuscript writing. 
Z.Y. and A.N. developed, configured, and executed the numerical model, generated the datasets, and performed post-processing. 
B.E. designed and implemented the machine learning model. 
J.S. and P.R. carried out model training and inference. 
J.S. organized and uploaded all data and codes to Zenodo. 
S.S. performed S-parameter extraction and analysis. 
S.T. performed the Floquet nonlinear analysis. 
Y.T. helped with the training and inference.
Y.Z. and J.-M.H. performed the 1D theoretical analysis and the supplementary material writing. 
All authors discussed the results, contributed to writing and revising the manuscript, and approved the final version of the paper.

\section{Competing interests}
The authors declare no competing interests.

\section{References}
\label{sec:references}
\vspace{10mm}
\bibliographystyle{unsrt}
\bibliography{magnonphoton}

\newpage

\setcounter{figure}{0}
    \captionsetup[figure]{labelfont={bf}, name={Extended Data Figure}, labelsep=period}

\begin{figure}
    \centering
    \includegraphics[angle=0, origin=c, width=0.8\textheight]{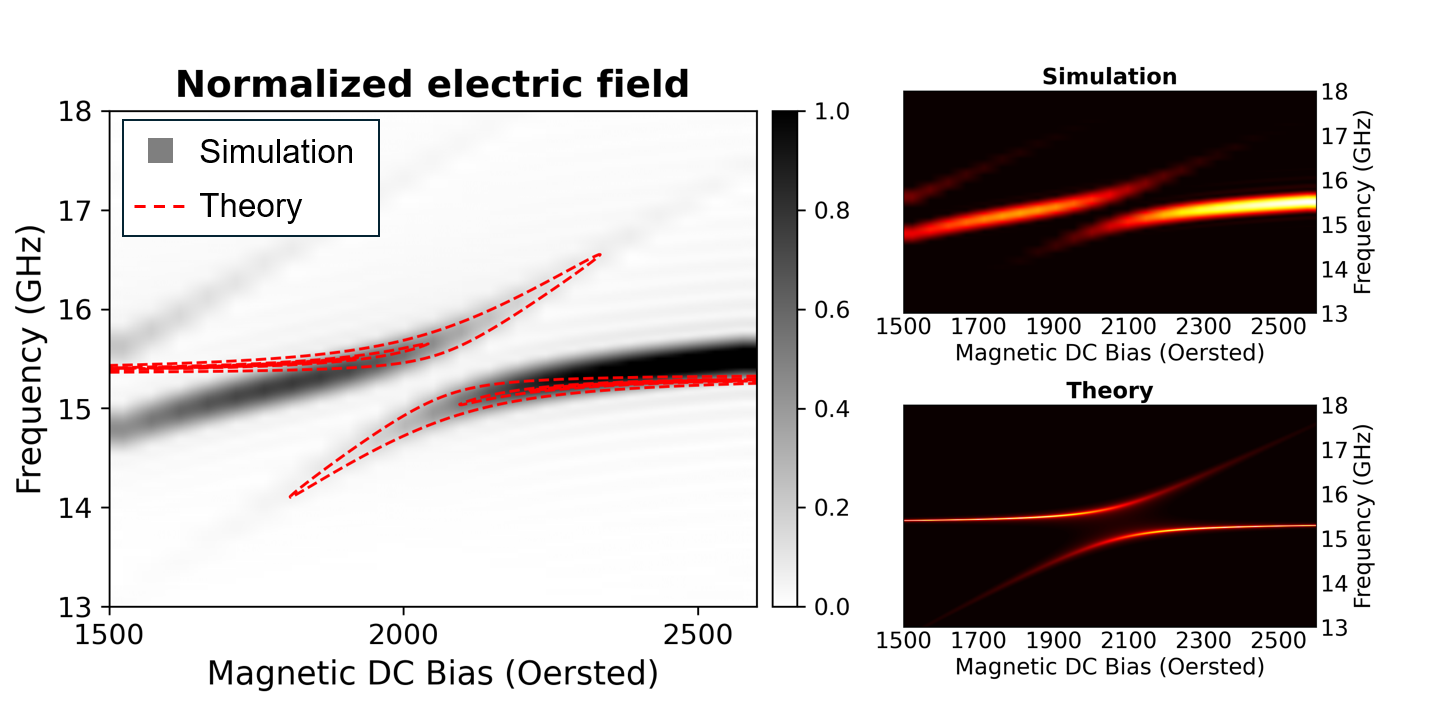}
    \caption{The theoretical analysis of the frequencies and linewidths of the hybrid polaritons agrees well with the numerical simulations. The right-hand figures show the color map of the simulated electric field (top) and the theoretically calculated electric field spectrum (bottom). In the left figure, the theoretical results are plotted as contour lines (red dashed) for clearer visualization.
    See \textbf{Supplementary Information} for details. }
    \label{fig:ext1-theory}
\end{figure}

\begin{figure}
    \centering
    \includegraphics[angle=90, origin=c, width=0.4\textheight]{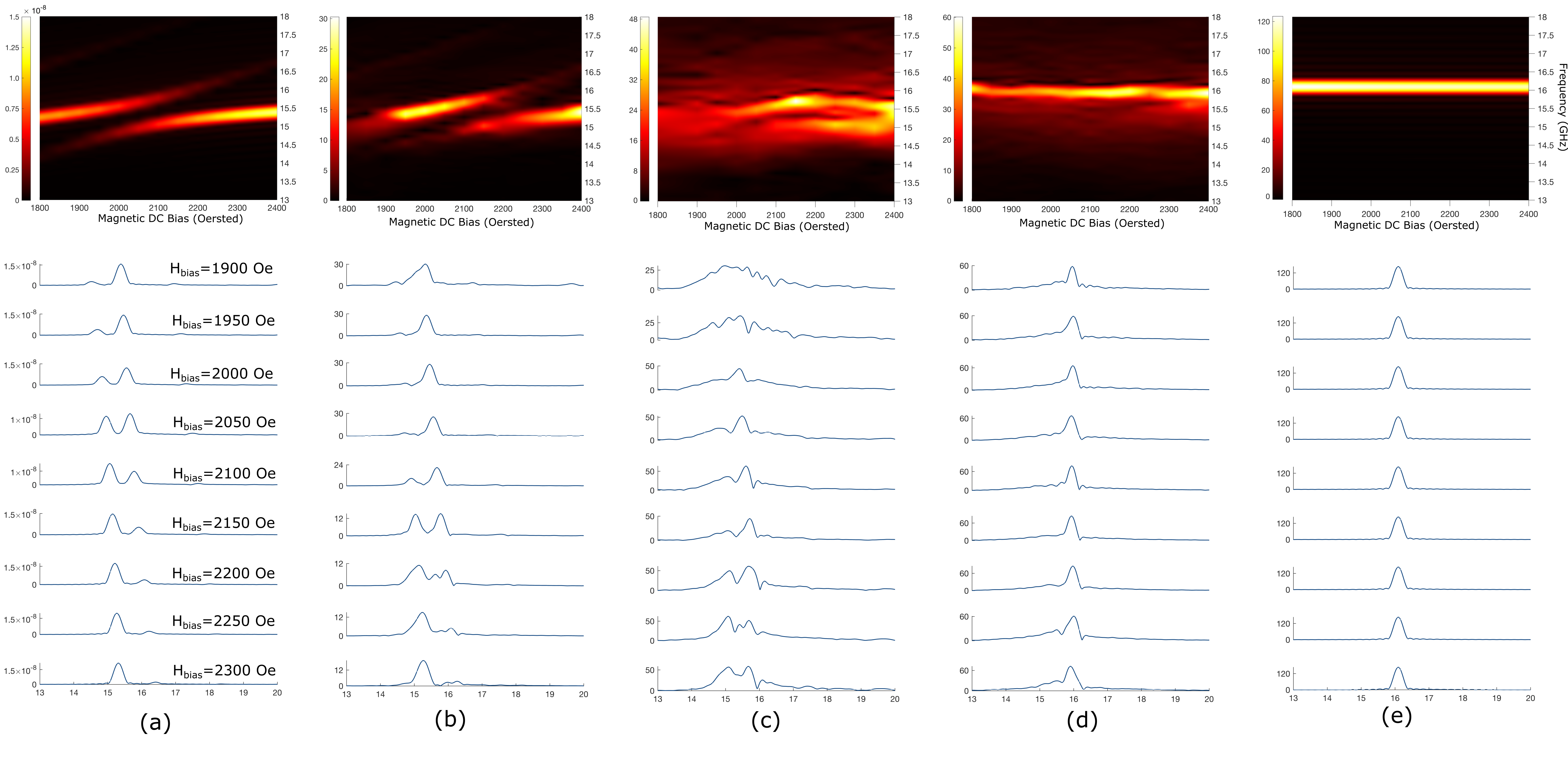}
    \caption{Electric-field spectra as the excitation intensity is increased. Stronger microwave magnetic fields suppress the magnon dynamics and leave only the cavity photon mode.  Such behaviour arises because, at high powers, nonlinear spin-wave interactions -- particularly three-magnon splitting -- induce a Suhl instability that transfers energy from the uniform ferromagnetic resonance to non-uniform spin waves; this saturation of the Kittel mode causes the magnon–photon anticrossing gap to close and yields a single photon-like resonance, a phenomenon sometimes described as the high-power cavity magnon-polaritons regime~\cite{nonlinear2023}. See \textbf{Supplementary Information} for details.  }
    \label{fig:ext2-nonlinear}
\end{figure}

\clearpage

\noindent \textbf{\large Supplementary Information: }

\vspace{0.2em}

\noindent \textbf{HPC-Driven Modeling with ML-Based Surrogates for Magnon-Photon Dynamics in Hybrid Quantum System}

\setcounter{section}{0}
\section{Analytical model of a 1D hybrid magnon--photon cavity}

Consider a 1D hybrid magnon-photon cavity where all physical quantities vary only along the $z$~axis; see Supplementary Fig.~\ref{fig:figS1-theory}. 

\setcounter{figure}{0}
    \captionsetup[figure]{labelfont={bf}, name={Supplementary Figure}, labelsep=period}
    
\begin{figure}
    \centering
    \includegraphics[width=\linewidth]{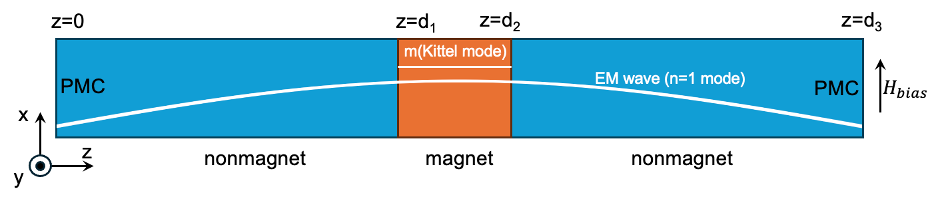}
    \caption{1D hybrid magnon-photon cavity. }
    \label{fig:figS1-theory}
\end{figure}

A bias magnetic field $\mathbf{H}^{\mathrm{bias}} = (H_x^{\mathrm{bias}},\, 0,\, 0)$ is applied along $+x$, stabilizing a spatially uniform magnetization, $\mathbf{m}(t{=}0)=\mathbf{m}^0=(1,0,0)$, along the same direction in the permalloy. 
Perfect magnetic conductor (PMC) boundary conditions, i.e.,\ $\mathbf{n}\times\mathbf{H}=0$ and $\mathbf{n}\cdot\mathbf{E}=0$ with $\mathbf{n}\parallel \hat{\mathbf z}$, confine a half-wavelength standing EM wave of angular frequency $\omega_{\mathrm{ph}}$. The photon frequency $\omega_{\mathrm{ph}}$ is set by the cavity length and does not vary with $H^{\mathrm{bias}}$.

Because the EM wavelength is much larger than the magnet thickness $d_2-d_1$, the EM fields are approximately uniform inside the magnet ($k_{\mathrm{ph}}\approx 0$). A microwave magnetic drive
\[
\mathbf{h}(t) = \left(h_x^0 e^{-\ii\omega t},\; h_y^0 e^{-\ii\omega t},\; h_z^0 e^{-\ii\omega t}\right)
\]
of angular frequency $\omega$ is applied to the magnet region and excites a spatially uniform (Kittel) magnon with zero wavevector $k_m{=}0$. When $\omega$ equals the Kittel resonance $\omega_m$ (FMR), the magnon amplitude peaks and the absorbed microwave power $P_{\mathrm{abs}}$ is maximized. The resonance $\omega_m$ is tunable with $H^{\mathrm{bias}}$. When $\omega_m$ is tuned close to $\omega_{\mathrm{ph}}$, the Kittel magnon and cavity photon hybridize into a magnon-polariton whose mode splitting is observable in the map $P_{\mathrm{abs}}(\omega,H^{\mathrm{bias}})$. Below we derive $P_{\mathrm{abs}}(\omega,H_x^{\mathrm{bias}})$ for fixed field direction ($+x$) and fixed drive amplitude:
\begin{equation}
P_{\mathrm{abs}}(\omega,H^{\mathrm{bias}})
\;\propto\;
\operatorname{Im}\!\left(\mathbf{h}^{\ast\,\mathsf T}\Delta\mathbf{m}\right)
=
\operatorname{Im}\!\left(\mathbf{h}^{\ast\,\mathsf T}\boldsymbol{\chi}\,\mathbf{h}\right)
=
\operatorname{Im}\!\left(h_i^0\,\chi_{ij}\,h_j^0\right),\quad i,j\in\{x,y,z\},
\tag{S1}
\label{eq:S1}
\end{equation}
where the (complex) magnetic susceptibility $\boldsymbol{\chi}$ is defined by $\Delta\mathbf{m}=\boldsymbol{\chi}\,\mathbf{h}$ with $h_i=h_i^0 e^{-\ii\omega t}$ and $h_i^\ast = h_i^0 e^{\ii\omega t}$. The tensor $\boldsymbol{\chi}$ follows from the linearized Landau--Lifshitz--Gilbert (LLG) equation and Maxwell's equations. The LLG equation reads
\begin{equation}
\frac{\partial \mathbf{m}}{\partial t}
=
-\gamma\,\mathbf{m}\times\left(\mathbf{H}^{\mathrm{eff}}+\mathbf{h}\right)
+\alpha\,\mathbf{m}\times\frac{\partial \mathbf{m}}{\partial t}.
\tag{S2}
\end{equation}
For an isotropic permalloy supporting only the Kittel mode, and neglecting magnon--phonon coupling, 
the effective field is
\[
\mathbf{H}^{\mathrm{eff}}=\mathbf{H}^{\mathrm{bias}}+\mathbf{H}^{\mathrm{EM}}+\mathbf{H}^{\mathrm{d}}+\mathbf{h},
\qquad
\mathbf{H}^{\mathrm{EM}}=(H_x^{\mathrm{EM}},\,H_y^{\mathrm{EM}},\,0),
\qquad
\mathbf{H}^{\mathrm{d}}=(0,\,0,\,-M_s m_z).
\]
In this 1D geometry $H_z^{\mathrm{EM}}(z,t)=0$; the time-varying $m_z(t)$ produces only a dynamic demagnetizing field $\Delta H_z^{\mathrm{d}}(t)=-M_s m_z(t)$. Let $\mathbf{m}=\mathbf{m}^0+\Delta\mathbf{m}(z,t)$, with
\[
\mathbf{m}=\Big(m_x^0+\Delta m_x^0 e^{-\ii\omega t},\;
\Delta m_y^0 e^{-\ii\omega t},\;
\Delta m_z^0 e^{-\ii\omega t}\Big).
\]
Keeping only linear terms yields
\begin{equation}
\begin{pmatrix}
\ii\omega & 0 & 0\\[2pt]
0 & \ii\omega & A_{23}\\[2pt]
0 & A_{32} & \ii\omega
\end{pmatrix}
\!
\begin{pmatrix}
\Delta m_x^0 e^{-\ii\omega t}\\[2pt]
\Delta m_y^0 e^{-\ii\omega t}\\[2pt]
\Delta m_z^0 e^{-\ii\omega t}
\end{pmatrix}
=
\gamma
\begin{pmatrix}
0\\[2pt]
- m_x^0\,h_z^0 e^{-\ii\omega t}\\[2pt]
\;\;m_x^0\,h_y^0 e^{-\ii\omega t}
\end{pmatrix}
+
\begin{pmatrix}
\Omega_x(z,t)\\[2pt]
\Omega_y(z,t)\\[2pt]
\Omega_z(z,t)
\end{pmatrix},
\tag{S3}
\label{eq:llg_linear}
\end{equation}
with
\begin{equation}
A_{23}=\ii\alpha\omega-\gamma H_x^{\mathrm{bias}}-\gamma M_s,
A_{32}=-\ii\alpha\omega+\gamma H_x^{\mathrm{bias}},
\tag{S4a}
\label{eq:A}
\end{equation}
\begin{equation}
\Omega_x(z,t)=0,\quad \Omega_y(z,t)=0,\quad \Omega_z(z,t)=\gamma\,H_y^{\mathrm{EM}}(z,t).
\tag{S4b}
\label{eq:Omega}
\end{equation}
The quantities $\Omega_i$ describe the bidirectional magnon--photon coupling (units of Hz). Introduce an auxiliary coupling matrix $\vartheta_{ij}$ via
\[
\Omega_i(z,t)=\vartheta_{ij}(z)\,\Delta m_j(t)=\vartheta_{ij}(z)\,\Delta m_j^0 e^{-\ii\omega t}.
\]
Then \eqref{eq:llg_linear} and \eqref{eq:Omega} lead to
\begin{equation}
\begin{pmatrix}
\Delta m_x^0 e^{-\ii\omega t}\\[2pt]
\Delta m_y^0 e^{-\ii\omega t}\\[2pt]
\Delta m_z^0 e^{-\ii\omega t}
\end{pmatrix}
=
\boldsymbol{\chi}
\begin{pmatrix}
h_x^0 e^{-\ii\omega t}\\[2pt]
h_y^0 e^{-\ii\omega t}\\[2pt]
h_z^0 e^{-\ii\omega t}
\end{pmatrix},
\tag{5a}
\label{eq:S5a}
\end{equation}
\begin{equation}
\boldsymbol{\chi}
=
\gamma
\begin{pmatrix}
\ii\omega & 0 & 0\\[2pt]
0 & \ii\omega & A_{23}\\[2pt]
-\vartheta_{zx} & A_{32}-\vartheta_{zy} & \ii\omega-\vartheta_{zz}
\end{pmatrix}^{\!-1}
\!
\begin{pmatrix}
0 & 0 & 0\\
0 & 0 & -1\\
0 & 1 & 0
\end{pmatrix}.
\tag{5b}
\label{eq:S5b}
\end{equation}
The unknowns in $\boldsymbol{\chi}$ are $\vartheta_{zi}$, which follow from
\[
\Omega_z(z,t)=\gamma H_y^{\mathrm{EM}}(z,t)=\vartheta_{zj}(z)\,\Delta m_j(t),
\]
once $H_y^{\mathrm{EM}}(z,t)$ is obtained from Maxwell's equations under PMC boundaries.

\paragraph{EM field solution (1D).}
With $\mathbf{H}^{\mathrm{EM}}=(H_x,H_y,0)$ and uniform $\mathbf{M}(t)=(M_x,M_y,M_z)$, Maxwell--Amp\`ere's law $\nabla\times\mathbf{H}=\mathbf{J}+\partial\mathbf{D}/\partial t$ with $\mathbf{J}=\sigma\mathbf{E}$ and $\mathbf{D}=\epsilon_0\epsilon_r\mathbf{E}$ yields
\begin{equation}
\nabla(\nabla\cdot\mathbf{H})-\nabla^2\mathbf{H}
=
\nabla\times\left(\sigma\mathbf{E}+\frac{\partial\mathbf{D}}{\partial t}\right).
\tag{6}
\end{equation}
Using $\mathbf{B}=\mu_0(\mathbf{H}+\mathbf{M})$, Faraday's law $\nabla\times\mathbf{E}=-\partial\mathbf{B}/\partial t$, and assuming plane-wave time dependence $e^{-\ii\omega t}$, the $x,y$ components satisfy
\begin{equation}
\frac{\partial^2 H_i(z)}{\partial z^2}+k^2 H_i(z)=-k^2 M_i,
\qquad
k^2=\epsilon_0\epsilon_r\mu_0\omega^2+\ii\sigma\mu_0\omega,
\qquad i\in\{x,y\}.
\tag{S7}
\end{equation}
In the magnet ($m$) and in the nonmagnetic cavity region ($c$), the 1D solutions are
\begin{subequations}
\label{eq:Hsolutions}
\begin{align}
H_i^{m}(z) &= H_i^{(m+)} e^{\ii k^m z}+H_i^{(m-)} e^{-\ii k^m z}-M_i,
\tag{S8a}\\
H_i^{c}(z) &= H_i^{(c+)} e^{\ii k^c z}+H_i^{(c-)} e^{-\ii k^c z},
\tag{S8b}
\end{align}
\end{subequations}
where $H_i^{(m\pm)}$ and $H_i^{(c\pm)}$ are the forward/backward wave amplitudes in magnet and cavity, respectively. From $\nabla\times\mathbf{H}=(\sigma-\ii\epsilon_0\epsilon_r\omega)\mathbf{E}$, one obtains
\begin{subequations}
\begin{align}
E_i^{m}(z) &= \big(\sigma^{m}-\ii\epsilon_0\epsilon_r^{m}\omega\big)^{-1}\!\left(-\frac{\partial H_j^{m}(z)}{\partial z}\right),\qquad i\neq j,\ i,j\in\{x,y\},
\tag{S9a}\\
E_i^{c}(z) &= \big(\sigma^{c}-\ii\epsilon_0\epsilon_r^{c}\omega\big)^{-1}\!\left(-\frac{\partial H_j^{c}(z)}{\partial z}\right),
\tag{S9b}
\end{align}
\end{subequations}
with PMC boundaries $H_i^{c}(z)=0$ at $z=0,d_3$ and EM continuity $H_i^{c}=H_i^{m}$, $E_i^{c}=E_i^{m}$ at $z=d_1,d_2$ ($i=x,y$). These conditions determine $H_i^{(c\pm)}$ and $H_i^{(m\pm)}$ in terms of $M_i$.

Consequently, $H_y^{\mathrm{EM}}(z,t)\equiv H_y^{m}(z)e^{-\ii\omega t}$. For the parameter set in Table~\ref{tab:params}, at resonance $\omega/2\pi=\omega_m/2\pi=\SI{14.17}{GHz}$ with $H_x^{\mathrm{bias}}=\SI{1800}{Oe}$, one finds
\begin{equation}
H_y^{\mathrm{EM}}(z,t)
=
\Big[
(424551-257055\,\ii)\,e^{\ii k^{m} z}
+
(424224+256900\,\ii)\,e^{-\ii k^{m} z}
\Big]\,
\Delta m_y^0\, e^{-\ii\omega t},
\quad d_1<z<d_2,
\tag{S10}
\label{eq:S10}
\end{equation}
with $k^{m}=(297.033+0.188\,\ii)\ \mathrm{m}^{-1}$, whose imaginary part encodes ohmic loss. From the definition of $\vartheta_{ij}$,
\begin{equation}
H_y^{\mathrm{EM}}(z,t)
=
\frac{1}{\gamma}\Big(\vartheta_{zx}(z)\,\Delta m_x^0+\vartheta_{zy}(z)\,\Delta m_y^0+\vartheta_{zz}(z)\,\Delta m_z^0\Big)\,e^{-\ii\omega t}.
\tag{S11}
\label{eq:S11}
\end{equation}
Comparing \eqref{eq:S10} with \eqref{eq:S11} shows $\vartheta_{zx}=\vartheta_{zz}=0$ and
\begin{equation}
\vartheta_{zy}(z)
=
(424551-257055\,\ii)\,e^{\ii k^{m} z}
+
(424224+256900\,\ii)\,e^{-\ii k^{m} z},
\qquad d_1<z<d_2.
\tag{S12}
\end{equation}
Because the EM wave is nearly uniform in the magnet,
\begin{equation}
H_y^{\mathrm{EM}}(t)
\approx
\langle H_y^{m}(z)\rangle e^{-\ii\omega t}
=
\frac{\langle \vartheta_{zy}(z)\rangle}{\gamma}\,\Delta m_y^0\,e^{-\ii\omega t}
=
\frac{1}{\gamma(d_2-d_1)}
\left(\int_{d_1}^{d_2}\vartheta_{zy}(z)\,dz\right)\Delta m_y^0 e^{-\ii\omega t}.
\tag{S13}
\end{equation}
For $H_x^{\mathrm{bias}}=\SI{1800}{Oe}$ and $\omega=\omega_m$, we obtain
$\langle \vartheta_{zy}\rangle = 4.966\times10^{9}+5.859\times10^{6}\,\ii\ \mathrm{Hz}$,
giving $\Omega_z(t)=(-2.222\times10^{8}+1.349\times10^{7}\,\ii)\,e^{-\ii\omega_m t}\ \mathrm{Hz}$ for
$\Delta m_y^0\approx -4.473\times10^{-2}+2.7684\times10^{-3}\,\ii$ (the imaginary parts capture EM loss and phase lag). Using $\langle \vartheta_{zy}\rangle$ in \eqref{eq:S5b} yields the full $\boldsymbol{\chi}$ and, via \eqref{eq:S1}, $P_{\mathrm{abs}}$.

\begin{table}
\centering
\caption{Parameters used in the analytical example.}
\label{tab:params}
\begin{tabular}{@{}ll@{}}
\toprule
Parameter & Value \\
\midrule
$\sigma^{m}$ & $10^{-3}\ \mathrm{S/m}$ \\
$\sigma^{c}$ & $10^{-3}\ \mathrm{S/m}$ \\
$\varepsilon_r^{m}$ & $7.1$ \\
$\varepsilon_r^{c}$ & $1$ \\
$M_s$ & $9.7\times10^{5}\ \mathrm{A/m}$ \\
$\gamma$ & $2.23\times10^{5}\ \mathrm{Hz\,m/A}$ \\
$\alpha$ & $0.003$ \\
$d_2-d_1$ & $7\times10^{-7}\ \mathrm{m}$ \\
$d_3$ & $3.67\times10^{-3}\ \mathrm{m}$ \\
$h_x^0$ & $0$ \\
$h_y^0$ & $10^{3}\ \mathrm{A/m}$ \\
$h_z^0$ & $0$ \\
\bottomrule
\end{tabular}
\end{table}

\vspace{1ex}

\section{Nonlinear Effect Analysis - Floquet Engineering.}
\label{subsec:methods-floquet}

For the nonlinear effect analysis, let us ignore the damping term in the nonlinear analysis since the $\alpha$ value is set to be very small ($\alpha = 0.003$ in our setup).  
At the center of the ferrite thin film, the EM magnetic field  primarily involves $H_x$ and $H_z$ components, as shown in Figure 1 in the main manuscript. 
$H_y$ is considered to be minimal since the fundamental TEM mode is predicted in the simulation. 
Therefore, at the center of the ferrite, $\Heff$ can be decomposed into three components:
\begin{equation}
\Heff = \hat{x}H_x + \hat{y}H_0 + \hat{z}H_z ,
\label{eq:Heff}
\end{equation}
where $H_0$ is the constant magnetic bias and $H_{x,z}$ are the time-dynamic EM field components.
In a stand-alone CPW resonator without any ferrite thin films, the resonant magnetic field can be simply expressed as $H_x = H_{x0} \cos(2\pi f_0 t) \sin(\pi y/y_0)$, where $H_{x0}$ is a constant magnitude.
With the ferrite thin film, the fundamental mode EM fields get distorted and hence excites an extra $H_z$ component, which shares similar spatial and temporal distributions as $H_x$, giving an expression of $H_z = H_{z0} \cos(2\pi f_0 t) \sin(\pi y/y_0)$, where $H_{z0}$ is a constant magnitude.
Note that the EM $H_y$ component is negligible, as shown by the simulation result, which agrees with the fundamental mode property. 
In our setup, $y_0 = \lambda/2$, thus in the center of the ferrite, where $y = \lambda/4$, $\sin(\pi y/y_0) = 1$, which simplifies the expression of $H_x$ and $H_z$ into 
\begin{equation}
H_x = H_{x0} \cos(2\pi f_0 t), \hspace{0.5cm} H_z = H_{z0} \cos(2\pi f_0 t) .
\label{eq:Hx_EM}
\end{equation}
The simulated $H_x$ and $H_z$ fields contain two resonant peaks, i.e., 
\begin{equation}
H_x = H_{x1} \cos(2\pi f_1 t) + H_{x2} \cos(2\pi f_2 t), \hspace{0.5cm} H_z = H_{z1} \cos(2\pi f_1 t) + H_{z2} \cos(2\pi f_2 t) .
\label{eq:Hx_EM2}
\end{equation}
For simplicity, let's start with just one resonance as indicated in Eq. (\ref{eq:Hx_EM}).

To perform Floquet analysis, let us scalarize the LLG equation:
\begin{equation}
\frac{\partial M_x}{\partial t} = \mu_0 \gamma [M_y H_z - M_z (H_y+H_0)], 
\label{eq:scalarLLGx}
\end{equation}
\begin{equation}
\frac{\partial M_y}{\partial t} = \mu_0 \gamma [M_z H_x - M_x H_z], 
\label{eq:scalarLLGy}
\end{equation}
\begin{equation}
\frac{\partial M_z}{\partial t} = \mu_0 \gamma [M_x (H_y+H_0) - M_y H_x] ,
\label{eq:scalarLLGz}
\end{equation}
where $M_x^2 + M_x^2 + M_x^2 = M_S^2$.
The terms $H_x$ and $H_z$ are shown in Eq. (\ref{eq:Hx_EM}); and $H_y = 0$.

Substituting Eq. (\ref{eq:Hx_EM}) into Eqs. (\ref{eq:scalarLLGx}), (\ref{eq:scalarLLGy}) and (\ref{eq:scalarLLGz}) reduces them to
\begin{equation}
\frac{\partial M_x}{\partial t} = \mu_0 \gamma [M_y H_{z0} \cos(2\pi f_0 t) - M_z H_0], \hspace{0.3cm}
\label{eq:scalarLLGx2}
\end{equation}
\begin{equation}
\frac{\partial M_y}{\partial t} = \mu_0 \gamma [M_z H_{x0} \cos(2\pi f_0 t) - M_x H_{z0} \cos(2\pi f_0 t)],
\label{eq:scalarLLGy2}
\end{equation}
\begin{equation}
\frac{\partial M_z}{\partial t} = \mu_0 \gamma [M_x H_0 - M_y H_{x0} \cos(2\pi f_0 t)] .
\label{eq:scalarLLGz2}
\end{equation}

In matrix form, the above equations can be rewritten as:
\begin{equation}
\frac{\partial}{\partial t}
\begin{bmatrix}
M_x \\
M_y \\
Mz 
\end{bmatrix}
= 
\begin{bmatrix}
0 & H_{z0} \cos(2\pi f_0 t) & -H_{0} \\
-H_{z0} \cos(2\pi f_0 t) & 0 & H_{x0} \cos(2\pi f_0 t) \\
H_0 & -H_{x0} \cos(2\pi f_0 t) & 0
\end{bmatrix}
\begin{bmatrix}
M_x \\
M_y \\
Mz 
\end{bmatrix} .
\end{equation}
Note that $H_{x0}$, $H_{z0}$ and $H_0$ are all constants, and $H_{x0}$ and $H_{z0}$ increase when we increase the input EM power. 

Alternatively, the LLG equation can be written in spherical coordinates (note that the $y$-direction is the polar axis): $M_x = M_S \sin(\theta) \cos(\phi)$, $M_y = M_S \cos(\theta)$, and $M_z = M_S \sin(\theta) \sin(\phi)$. 
It follows that:
\begin{equation}
\frac{\partial [\sin(\theta) \cos(\phi)]}{\partial t} = \mu_0 \gamma [ \cos(\theta) H_{z0} \cos(2\pi f_0 t) -  \sin(\theta) \sin(\phi) H_0], \hspace{0.3cm}
\label{eq:scalarLLGx3}
\end{equation}
\begin{equation}
\frac{\partial [\cos(\theta)]}{\partial t} = \mu_0 \gamma [ \sin(\theta) \sin(\phi) H_{x0} \cos(2\pi f_0 t) - \sin(\theta) \cos(\phi) H_{z0} \cos(2\pi f_0 t)],
\label{eq:scalarLLGy3}
\end{equation}
\begin{equation}
\frac{\partial [\sin(\theta) \sin(\phi)]}{\partial t} = \mu_0 \gamma [\sin(\theta) \cos(\phi) H_0 - \cos(\theta) H_{x0} \cos(2\pi f_0 t)] .
\label{eq:scalarLLGz3}
\end{equation}
Further simplifying this, we obtain:
\begin{equation}
-\sin(\theta) \sin(\phi) \frac{\partial \phi}{\partial t} + \cos(\theta) \cos(\phi) \frac{\partial \theta}{\partial t} =  \mu_0 \gamma [ \cos(\theta) H_{z0} \cos(2\pi f_0 t) -  \sin(\theta) \sin(\phi) H_0], \hspace{0.3cm}
\label{eq:scalarLLGx4}
\end{equation}
\begin{equation}
-\sin (\theta) \frac{\partial \theta}{\partial t} =  \mu_0 \gamma [ \sin(\theta) \sin(\phi) H_{x0} \cos(2\pi f_0 t) - \sin(\theta) \cos(\phi) H_{z0} \cos(2\pi f_0 t)],
\label{eq:scalarLLGy4}
\end{equation}
\begin{equation}
\sin(\theta) \cos(\phi) \frac{\partial \phi}{\partial t} + \cos(\theta) \sin(\phi) \frac{\partial \theta}{\partial t} = \mu_0 \gamma [\sin(\theta) \cos(\phi) H_0 - \cos(\theta) H_{x0} \cos(2\pi f_0 t)] .
\label{eq:scalarLLGz4}
\end{equation}

\end{document}